# Navigating the skill diversity frontier: How skill complexity explains worker resilience

Mar Carpanelli[a,b], Jedrzej Duszynski[a] & Fabian Stephany [a,c,d]

a) Oxford Internet Institute, University of Oxford, UK, b) Economic Graph Research Institute, LinkedIn Corporation, Sunnyvale, California, USA c) Institute for New Economic Thinking, Oxford Martin School, d) Bruegel, Brussels, Belgium.

fabian.stephany@oii.ox.ac.uk

# Abstract

As artificial intelligence transforms labor markets, understanding what makes workers adaptable has become increasingly important. Existing approaches typically characterize human capital using occupations, educational credentials, or predefined skill taxonomies, providing limited insight into how the structure of workers' skill portfolios shapes resilience to technological change. We develop an agnostic network based framework that reconstructs the hierarchy and diversity of skills directly from observed patterns of skill co occurrence. Using longitudinal data on 2.4 million United States workers and 16,753 distinct skills from LinkedIn, we introduce three complementary measures of skill complexity: specialisation, capturing productive depth; diversity, capturing adaptive breadth; and the diversity frontier, measuring the highest attainable diversity conditional on a worker's level of specialisation. We show that these dimensions predict distinct career outcomes. Specialisation is most strongly associated with sorting into higher wage occupations, whereas diversity is associated with broader skill accumulation and occupational mobility. Workers closest to the diversity frontier are significantly more likely to acquire new skills, receive promotions, transition into occupations with lower exposure to automation than workers with comparable levels of specialisation but narrower skill portfolios. These findings distinguish productive from adaptive capital and demonstrate that workers' adaptive capacity depends not simply on possessing specialised expertise or broad capabilities, but on combining both. More broadly, our framework provides a data driven approach for measuring workforce resilience and identifying reskilling pathways, offering new tools for understanding human capital in rapidly changing labor markets.

# Introduction

How do workers remain resilient to technological change? This question has become increasingly urgent as artificial intelligence (AI) reshapes the organisation of work. Recent technological change does not merely automate routine tasks; it increasingly complements, recombines, and reconfigures human capabilities across occupations. A central insight from the literature on task-biased and skill-biased technological change is that disruption occurs at the tasks and skills level, rather than at the occupation level (Autor, Levy and Murnane, 2003; Acemoglu and Autor, 2011). This is because occupations are bundles of tasks that can be substituted, complemented, or augmented by technology. Workers are therefore increasingly required to deprioritise tasks and skills that become automatable, while developing capabilities that remain valuable under new technological conditions. In this sense, resilience depends not only on whether a worker's current occupation is exposed to automation, but on whether their skills allow them to adapt (Manning and Aguirre, 2026).

Prior research has conceptualised worker resilience through resistance to displacement, recovery after job loss, and adaptive capacity: the ability to transition into new tasks, roles, or skill domains (Autor, Dorn and Hanson, 2013; Acemoglu and Restrepo, 2019; Dauth et al., 2021; Huckfeldt, 2022). In the context of automation, adaptive capacity becomes especially important. Workers remain competitive not because their existing tasks are fully insulated from automation, but because they can acquire, combine, and redeploy skills as technologies change (Deming, 2017; Pedota, Grilli and Piscitello, 2023; Stephany and Teutloff, 2024; Bone, González Ehlinger and Stephany, 2025; Mäkelä et al., 2025).

We use resilience to refer to workers' ability to maintain labour market position and adapt through learning and mobility. Our main theoretical focus is adaptive capacity, one dimension of resilience. Yet the micro-level determinants of adaptability remain insufficiently understood. Existing models of technological change have shifted attention from occupations to tasks and skills, but they tell us less about the adaptability potential embedded in workers' skill sets themselves. In knowledge-intensive labour markets, where production depends on cumulative and interdependent capabilities, the relevant margin of adjustment may lie not only in what workers do, but in how their skills are structured (Neffke and Henning, 2013; Frank *et al.*, 2019; Lee, Jeong and Lee, 2025).

Our paper argues that skill complexity offers a systemic perspective on worker adaptability and resilience. We define skill complexity as the structure of workers' skill portfolios within a broader hierarchy and domain system: whether workers' skills are concentrated in more specialised capabilities, and whether those skills span multiple domains. This allows us to decompose workers' human capital into two conceptually distinct components. The first is productive capital: specialised expertise that may be valuable within existing labour market structures. The second is adaptive capital: diversity across skill domains that may provide flexibility, learning pathways, and mobility under changing technological conditions.

Workers with highly specialised skill portfolios may benefit from immediate labour market rewards, particularly if specialised skills signal scarce or differentiated expertise. However, specialisation alone may also narrow the set of domains into which workers can move. Diverse skill portfolios, by contrast, may support adaptation by exposing workers to a broader set of domains and increasing the number of pathways through which they can acquire new skills, move across roles, or respond to changing task demands (Lazear, 2009; Gathmann and Schönberg, 2010). The key question is therefore not simply whether specialists or generalists are more resilient, but whether productive capital and adaptive capital operate independently or reinforce each other. Put differently, resilience may depend not on maximising specialisation or diversity in isolation, but on how much diversity workers achieve relative to their degree of specialisation. We aim to test this confining element of specialisation on diversity with a novel concept called "diversity frontier".

We test these ideas using large-scale LinkedIn data on workers' skills and subsequent labour market outcomes. We examine whether pre-existing skill portfolios predict multiple dimensions of resilience: labour market rewards, measured through sorting into higher-wage occupations and promotions; flexibility, measured through role transitions into occupations with lower automation exposure; and technological adaptation, measured through skill acquisition and the adoption of AI-relevant skills. Our empirical strategy estimates models that include three network-derived measures of skill portfolio structure: specialisation, diversity, and frontier position, that is a worker's degree of skill diversity relative to their level of specialisation. This allows us to distinguish productive depth, adaptive breadth, and the extent to which workers achieve greater diversity than peers with comparable levels of specialisation. Throughout, we interpret the results as evidence that the structure of pre-existing skill portfolios predicts different forms of subsequent labour market adaptation, rather than as causal estimates of the effects of particular skill configurations.

This paper makes three primary contributions to our understanding of labour market dynamics amid technological change. First, it introduces measures of skill complexity derived from agnostic metrics of skill diversity and specialisation to explain workers' resilience. Second, it decomposes human capital into productive and adaptive capital, showing that these dimensions correspond to different labour market outcomes. Third, it shows that adaptive outcomes are most closely associated not with specialisation or diversity in isolation, but with frontier position: greater cross-domain breadth relative to workers with comparable levels of specialised depth. Together, these contributions show how the structure of workers' skill portfolios shapes different forms of resilience to technological change.

# Background

## Technology reshuffles skills

Technological change has long been understood as a key driver of labour market transformation. In task-based models, workers are endowed with skills, while jobs consist of bundles of tasks that can be substituted or complemented by technology (Acemoglu and Autor, 2011). Early research on routine-biased technological change showed that automation disproportionately affected routine, codifiable tasks, contributing to job polarisation and wage inequality (Autor, Levy and Murnane, 2003; Goos, Manning and Salomons, 2014).

Recent advances in artificial intelligence extend this process beyond routine substitution. AI increasingly augments, recombines, and reorganises tasks across the skill distribution, meaning that technological disruption operates less at the level of entire occupations and more at the level of tasks and skills (Autor, 2015; Acemoglu and Restrepo, 2018, 2019). Worker resilience therefore depends not only on whether current tasks are automatable, but on whether workers can reallocate existing skills or acquire new ones in response to changing demand (Gathmann and Schönberg, 2010; Neffke, Otto and Weyh, 2017; Huckfeldt, 2022).

A further development in this literature shifts attention from individual skills to relationships between skills. Research on skill relatedness, complementarity, and skill networks shows that skills are not independent attributes, but interdependent capabilities that shape workers' mobility and adaptation pathways (Neffke and Henning, 2013; Alabdulkareem *et al.*, 2018; Frank *et al.*, 2019; Stephany and Teutloff, 2024; Hosseinioun *et al.*, 2025; Lee, Jeong and Lee, 2025; Mäkelä et al., 2025). Transitions are easier when new skills are related to existing ones, while movement across unrelated domains requires more costly investment. However, much of this work focuses on individual skills or pairwise relationships between skills. To understand adaptability more fully, we need to examine how workers' skills are positioned within the broader hierarchy and systemic structure of the entire system of skills.

## Skill complexity refines human capital

Traditional measures of human capital, such as education or occupation, provide limited insight into how workers' capabilities are organised and combined. Even when skills are observed directly, they are often treated as lists of attributes rather than structured portfolios. Recent work in economic complexity provides a different perspective, conceptualising skills as part of interconnected systems in which capabilities vary in their position, relatedness, and dependency structure (Hidalgo and Hausmann, 2009; Tacchella *et al.*, 2012; Frank *et al.*, 2019; Hosseinioun *et al.*, 2025; Lee, Jeong and Lee, 2025). Building on this perspective, we conceptualise skill complexity as the structure of a worker's skill portfolio within the broader skill system.

Our contribution is to interpret skill complexity as a decomposition of human capital into two dimensions: productive capital and adaptive capital. Productive capital refers to the market value associated with specialised expertise. Adaptive capital refers to the flexibility generated by spanning multiple skill domains. Furthermore, we postulate that both dimensions have a non-linear relationship as the potential diversity of each worker is constrained by their degree of specialisation. Potential skill diversity is bounded from both directions: highly specialized workers are constrained by the cumulative investments required to acquire deep domain expertise, whereas highly generalist workers are constrained because broad foundational capabilities increasingly converge across occupations, limiting the number of distinct knowledge domains available for further diversification.

The first dimension is hierarchical position. Skills occupy different positions within a dependency hierarchy, ranging from broadly applicable foundation skills to more specialised downstream capabilities. For example, statistical skills may underpin machine learning, which may in turn enable further specialisation in generative AI models. Workers whose skills are concentrated in more downstream positions possess deeper and more specialised expertise. Downstream skills are interpreted as productive capital because they are less foundational, more domain-specific, and typically require the accumulation of upstream capabilities, making them stronger signals of differentiated expertise. We therefore expect downstream concentration to be associated with sorting into higher-wage occupations. At the same time, specialised skills may be less transferable when demand shifts, because their value is often tied to narrower domains of application (Becker, 1962; Spence, 1973; Lazear, 2009; Gathmann and Schönberg, 2010; Hanushek *et al.*, 2017; Eggenberger, Janssen and Backes-Gellner, 2022).

The second dimension is domain diversity. Workers may concentrate their skills within a single domain or combine capabilities from multiple domains of the skill network. We interpret diversity as adaptive capital: the breadth of knowledge areas that workers can draw on when learning new skills, moving across roles, or responding to changing labour market demands. This argument draws on theories of recombinant growth, which show how new capabilities emerge from novel combinations of existing knowledge components (Weitzman, 1998; Fleming, 2001). Related work on skill relatedness, brokerage, variety, and diversity similarly suggests that exposure to multiple knowledge domains expands transition pathways and supports flexibility (Burt, 2004; Frenken, Van Oort and Verburg, 2007; Stirling, 2007; Neffke and Henning, 2013; Neffke, Otto and Weyh, 2017; Huckfeldt, 2022).

## Specialisation confines diversity

Lastly, and most importantly, we argue that specialization and diversity should not be viewed as independent dimensions of skill complexity, but as jointly constrained characteristics of workers' skill portfolios. Recent work increasingly recognizes that human capital is organized as a nested hierarchy in which specialized capabilities build upon more fundamental prerequisite skills, implying that workers cannot freely combine arbitrary skills but instead accumulate them along structured developmental paths (Hosseinioun et al., 2025; Dorn et al., 2024). Building on this perspective, we argue that the degree of specialization shapes the feasible level of diversity, giving rise to what we term the diversity frontier: an inverse U-shaped relationship between specialization and the maximum attainable diversity.

Consider, for example, general managers and radiologists. General managers typically possess broad, transferable capabilities in leadership, communication, and strategic planning. While some broaden their portfolios through expertise in areas such as digital transformation or finance, their opportunities for further diversification are ultimately limited because additional skills increasingly draw on the same set of foundational managerial capabilities. Radiologists, by contrast, possess much deeper domain-specific expertise and therefore, on average, command higher labour market rewards. Although some diversify into areas such as AI-assisted diagnostics, research, or hospital management, the substantial investments required to acquire and maintain specialized expertise limit the extent to which they can broaden their skill portfolios. Consequently, both highly generalist and highly specialized workers face constraints on attainable skill diversity, albeit for different reasons: generalists because broad capabilities increasingly converge, and specialists because deep expertise requires cumulative investments within a single domain.

As a result, the greatest scope for skill diversity emerges at intermediate levels of specialization. We refer to the upper envelope of attainable diversity as the diversity frontier. Workers located on this frontier possess the greatest diversity feasible for their degree of specialization. Rather than maximizing breadth or depth in isolation, these workers optimally combine domain-specific expertise with cross-domain capabilities, making them particularly well positioned to acquire new skills, redeploy existing human capital across occupations, and adapt to technological change. We therefore expect these workers to be best positioned to acquire new skills, adapt to technological change, and transition across occupations.

This reasoning results in the following hypotheses:

**Hypothesis 1: Specialisation and rewards.** Workers with specialised skill portfolios are more likely to receive traditional labour market rewards, such as sorting into higher-wage occupations and being promoted.

**Hypothesis 2: Diversity and adaptability.** Workers with more diverse skill portfolios are more likely to display adaptive outcomes, including broader skill accumulation, occupational mobility into lower-exposure occupations, and AI-skill adoption.

**Hypothesis 3: Diversity frontier and adaptive capacity.** Workers located on or close to the diversity frontier are more likely to display significant adaptive outcomes. We expect frontier positions to be associated primarily with learning, occupational mobility, transitions into lower-exposure roles, and skill adoption.

# Methods

## Data

We use longitudinal data on workers' skill portfolios from LinkedIn, the world's largest professional networking platform, with more than 1.3 billion users as of May 2026. The platform contains detailed self-reported information on workers' skills, employment histories, and career progression, organised within structured taxonomies linked to standard occupational and industry classifications. Our analysis focuses on a random sample of approximately 2.4 million US workers who listed at least five skills prior to the diffusion of generative AI (pre-2022). The data are structured as an individual-year panel, allowing us to observe workers' skill portfolios and career trajectories over time. To reduce concerns about reverse causality, all explanatory variables are measured between 2020 and 2022, while outcomes are measured between 2023 and 2025.

# Metrics

## Skill network and independent variables

We construct a directed skill network from observed co-occurrence patterns across worker profiles. We first identify statistically meaningful skill associations by comparing observed co-occurrence to a hypergeometric null model. We then infer directionality from asymmetric conditional probabilities: if workers with a more specialised skill are more likely to also hold a broader skill than the reverse, we interpret the broader skill as upstream and draw a directed edge from the upstream skill to the downstream skill. Weak directional edges are pruned to recover a sparse and interpretable network. Appendix A2 provides full details on network construction, threshold selection, and robustness. We identify broader skill domains using Louvain community detection on the undirected projection of the skill network. These domains form the basis for measuring the diversity of workers' skill portfolios (Figure 1B).

We derive three main worker-level measures. First, specialisation captures whether a worker's portfolio is concentrated in downstream skills. At the skill level, we calculate local reaching centrality (LRC), defined as the share of skills reachable through directed paths. Skills with higher LRC occupy more upstream positions and are interpreted as more general, while skills with lower LRC occupy more downstream positions and are interpreted as more specialised. We normalise LRC within skill domains and define worker-level specialisation as the average of $1 - LRC_{norm}$ across all skills in a worker's portfolio. Higher values indicate more specialised portfolios.

Second, diversity captures the breadth of domains represented in a worker's skill portfolio. We measure diversity using Shannon-based Hill diversity across Louvain skill domains. This measure captures the effective number of skill domains represented in a portfolio, accounting for both the number of domains and the balance of skills across them. Higher values indicate broader cross-domain skill portfolios. Because workers with more listed skills mechanically have more opportunities to span multiple domains, all regressions control for total portfolio size. The coefficient on Hill diversity therefore captures breadth across domains adjusted for the number of skills listed.

Third, frontier position captures how much diversity a worker achieves relative to others with comparable levels of specialisation. This measure is motivated by the fact that diversity is constrained by specialisation via costs (specialists) and convergence (generalist). To construct the diversity frontier score, we first rank workers by their specialisation score and partition them into 100 equally sized groups. Within each specialisation group, we identify the maximum observed Hill diversity score. Each worker's Hill diversity is then divided by this group-specific maximum. The resulting measure ranges from 0 to 1 and captures how close a worker is to the observed diversity frontier among similarly specialised workers. Higher values indicate greater cross-domain diversity relative to workers with comparable levels of specialisation.

### Dependent variables

We consider six primary outcomes grouped into three domains of worker resilience. First, reward outcomes include: (i) promotion, defined as upward career progression into a higher-seniority role; and (ii) occupation-level wage sorting,[1] measured as the logarithm of the median national wage for each worker's stated occupation, ingested from O*NET.

Second, adaptability outcomes include: (iii) lateral transition, defined as movement across roles at a similar seniority level; and (iv) lateral transition into a role with lower AI exposure. We compute AI exposure at the occupation level based on the overlap between an occupation's skill portfolio and tasks identified as replicable by AI. Following the LinkedIn Economic Graph methodology (Carpanelli *et al.*, 2024), exposure is calculated by evaluating the potential for AI to perform specific occupational tasks based on the most prevalent skills of an occupation across LinkedIn data. To avoid mechanical overlap between worker-level skill portfolios and exposure outcomes, our primary lower-exposure transition measure uses exposure scores that are fixed within occupations and do not depend on the focal worker's own skill portfolio.

Third, technological adaptation is captured by (v) AI skill adoption, defined as the acquisition of skills related to the use of artificial intelligence tools, such as prompting; and by (vi) total skill acquisition, measured as the total number of new skills added. Because skill additions may reflect profile-updating behaviour as well as actual learning, all models control for prior profile update activity, and robustness specifications examine whether results hold among workers with comparable baseline activity levels.

## Empirical strategy

We examine whether pre-existing skill portfolio structure predicts subsequent labour market outcomes. All explanatory variables are measured before the outcome period. Our main specifications include three standardised skill complexity measures: specialisation, measured as average inverse normalised local reaching centrality; diversity, measured as Hill diversity across skill domains; and the diversity frontier score, which captures how close a worker is to the maximum observed diversity among workers with comparable levels of specialisation.

For each primary outcome, we estimate parallel fully controlled models that introduce each skill complexity measure separately, along with individual controls, topology controls, and fixed effects. This design allows us to compare how three dimensions of skill portfolio structure relate to outcomes: depth, captured by specialisation; breadth, captured by diversity; and breadth relative to depth, captured by the diversity frontier score.

---

[1] Because individual wages are not observable in LinkedIn data, this measure captures occupational sorting (whether workers with certain skill configurations sort into higher-paying occupations) rather than within-occupation wage variation. The substantive interpretation is therefore about where workers land in the occupational wage distribution, not about earnings premiums conditional on occupation.

Let $SC_{i,t-1}$ denote one of the three standardised skill complexity measures: specialisation, diversity, or frontier position. Our baseline specification for each measure is:

$$Y_{i,t} = \beta \ SC_{i,t-1} + \gamma X_{i,t-1} + \delta_{j(i)} + \varepsilon_i.$$

Model 1 includes base controls, topology controls, and fixed effects. In Models 2–4, $SC_{i,t-1}$ is replaced respectively by specialisation, diversity, and the diversity frontier score.

The vector $X_{i,t-1}$ includes individual-level controls: an indicator for holding multiple jobs, profile activity measured as the number of profile updates, network size measured as the number of connections, age, gender, education, and the total number of listed skills. We include fixed effects $\delta_{j(i)}$ for industry, company size, and functional area, capturing systematic differences across sectors, organisational contexts, and occupational groupings. Standard errors are clustered at the member level.

# Results

## Skill network and worker-level portfolio distributions

We model human capital as a structured ecosystem of capabilities by mapping the skill portfolios of approximately 2.4 million US workers into a directed skill network. Using the skills listed on LinkedIn profiles, we construct a network that allows us to measure both the hierarchical position of skills and the distribution of skills across domains. The resulting network contains 16,753 nodes, representing unique skills, and 556,206 directed edges, representing skill relationships. Louvain community detection identifies 12 broad skill domains.

Figure 1 visualises the network construction procedure, the resulting skill network, and the distribution of local reaching centrality across skill domains. The network is highly clustered but interconnected. High-centrality skills occupy more upstream positions and are embedded near the core of the network, while lower-centrality skills occupy more downstream and specialised positions. The distribution of local reaching centrality is highly skewed and zero-inflated, with a large share of downstream skills and a small number of highly central upstream skills. This structure provides the basis for measuring worker-level specialisation.

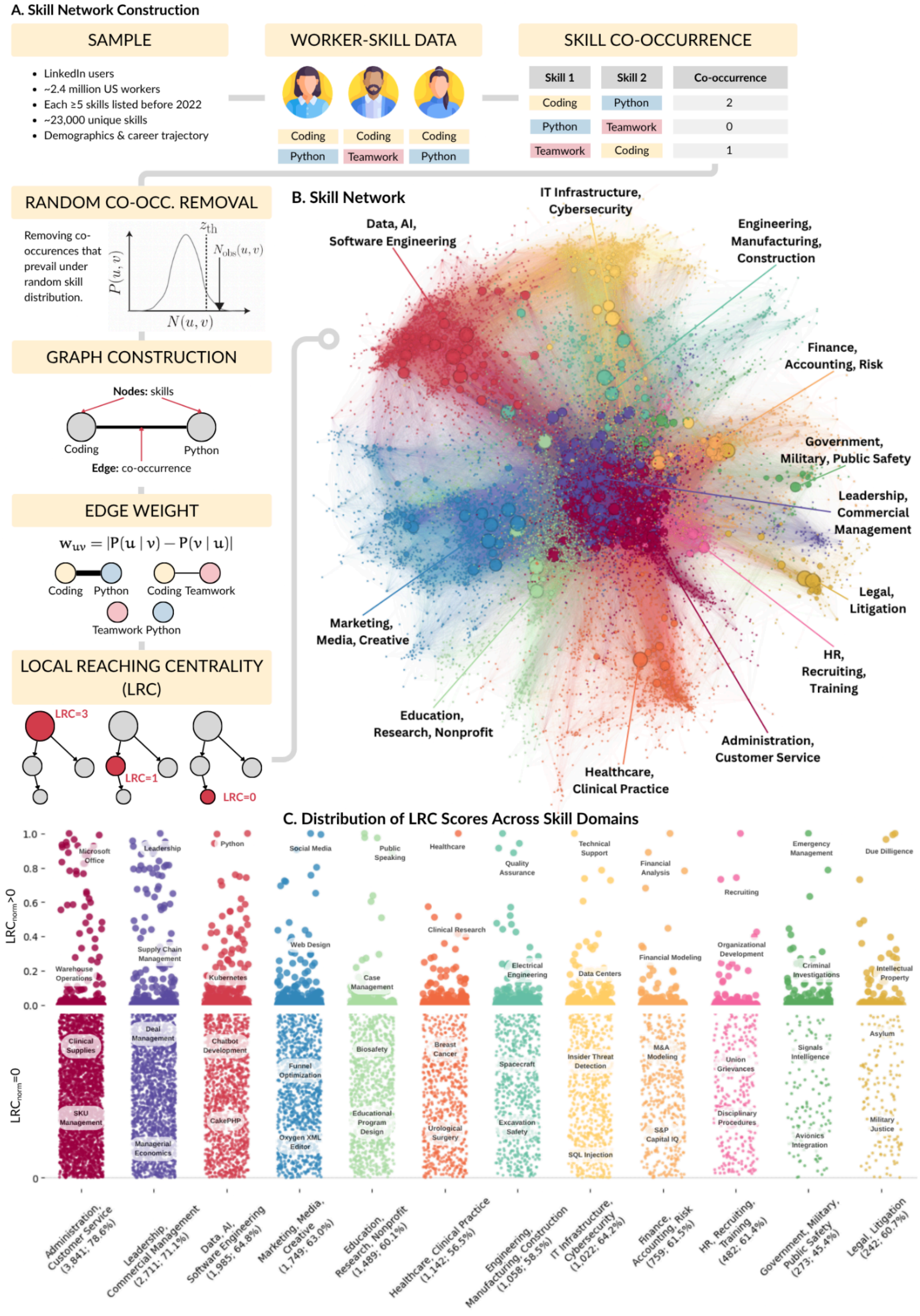
A. Skill Network Construction
SAMPLE
LinkedIn users
~2.4 million US workers
Each ≥5 skills listed before 2022
~23,000 unique skills
Demographics & career trajectory
WORKER-SKILL DATA
SKILL CO-OCCURRENCE
RANDOM CO-OCC. REMOVAL
Removing co-occurences that prevail under random skill distribution.
GRAPH CONSTRUCTION
Nodes: skills
Edge: co-occurrence
EDGE WEIGHT
$w_{uv} = |P(u \mid v) - P(v \mid u)|$
LOCAL REACHING CENTRALITY (LRC)
B. Skill Network
C. Distribution of LRC Scores Across Skill Domains
Skill Domain (No. Skills, Share of Downstream Skills)

**Figure 1. Skill network construction, structure, and hierarchical position of skills**
Panel A illustrates the construction of the skill network using LinkedIn data from approximately 2.4 million US workers, where nodes represent skills and edges capture relationships between skills within worker profiles. Edge weights reflect asymmetric co-occurrence probabilities after removing random associations. Panel B visualises the resulting network, with node size proportional to local reaching centrality, capturing the extent to which a skill provides access to downstream skills in the network. Colours indicate communities detected via the Louvain algorithm, corresponding to broad skill domains. Panel C shows the distribution of normalised local reaching centrality scores across skill domains, with examples of skills across the distribution.

Skill clusters vary considerably in size, with broad domains such as Administration and Customer Service, Leadership and Commercial Management, Data, AI and Software Engineering, and Marketing, Media and Creative containing the largest numbers of skills, whereas more specialized domains—including Finance, Accounting and Risk, HR and Recruiting, Government, Military and Public Safety, and Legal and Litigation—are comparatively smaller. Despite these differences in size, the share of terminal (downstream) skills with a local reaching centrality of zero is remarkably similar across clusters, averaging around 60%. Administration and Customer Service, together with Leadership and Commercial Management, are notable exceptions, exhibiting terminal skill shares exceeding 70%, reflecting a larger proportion of highly specialized downstream capabilities. For more details on skill cluster statistics, see Appendix A3.

Figure 2 summarises how portfolio measures vary across demographic groups. The first two columns show the distributions of specialisation and diversity, while the third column shows how mean diversity varies across the specialisation distribution. These patterns are consistent with recent evidence that human capital is accumulated through nested skill dependencies, whereby specialised capabilities build on increasingly general prerequisite skills, resulting in structured rather than arbitrary skill portfolios (Hosseinioun et al., 2025). They also accord with emerging evidence that multidimensional skill portfolios evolve systematically with education and work experience, with more educated and experienced workers accumulating larger shares of occupation specific and managerial skills (Dorn et al., 2024).

Specialisation shows clearer demographic stratification than diversity. Men are shifted toward higher specialisation than women, younger cohorts are more concentrated at higher levels of specialisation than older cohorts, and workers with higher educational attainment are shifted toward more specialised portfolios. These patterns are consistent with theories of life cycle human capital accumulation, whereby workers progressively acquire specialised capabilities through education and experience, although learning opportunities and the composition of skill portfolios evolve across the career (Ma et al., 2026). Diversity varies less sharply across groups, although younger cohorts and workers with bachelor's or graduate degrees tend to be concentrated around moderate to high levels of diversity. This is consistent with evidence that higher education broadens access to complementary knowledge domains before careers become increasingly specialised, while younger workers continue to explore a wider set of adjacent capabilities during early career development (Dorn et al., 2024; Desjardins & Huang, 2025).

The third column illustrates that diversity is also systematically related to specialisation, as hypothesised. Mean diversity rises from low to moderate levels of specialisation before declining among the most specialised workers, consistent with the notion that deeper expertise increasingly channels skill accumulation along narrower developmental pathways (Hosseinioun et al., 2025). This pattern appears both overall and within demographic subgroups, suggesting that skill complexity is structured jointly by nested skill dependencies, educational investments, occupational sorting, and career stage.

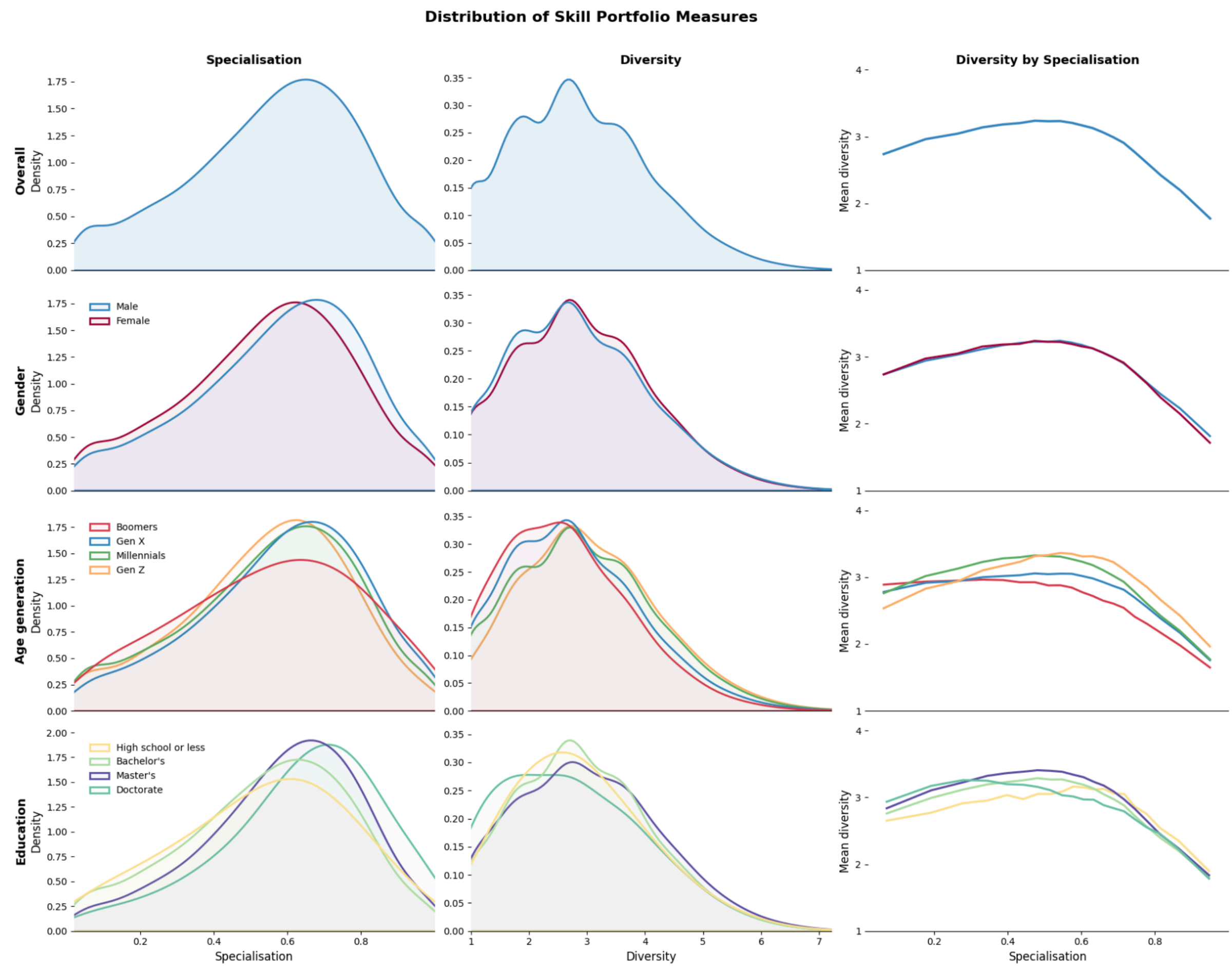

**Figure 2. Skill complexity by age, gender and education**
The first two columns show the 2025 distributions of specialisation and diversity. The third column plots mean diversity at each level of specialisation, illustrating how breadth varies across the depth distribution. Rows show the overall pattern and disaggregations by gender, age generation, and education.

## Skill complexity and career outcomes

We next test the hypotheses derived from our theory of productive and adaptive capital. Figure 3A maps workers across the skill portfolio space, with darker areas indicating higher outcome probabilities. Consistent with Hypothesis 1, sorting into higher wage occupations is concentrated among workers with more specialised portfolios, supporting the view that specialised skills constitute productive capital. In contrast, and in line with Hypothesis 2, skill acquisition and occupational mobility, including lateral transitions and moves into occupations with lower AI exposure, are increasingly concentrated among workers with more diverse skill portfolios. AI skill adoption is most prevalent near the diversity frontier, suggesting that adapting to emerging technologies requires both specialised expertise and broad capabilities spanning multiple knowledge domains.

Figure 3B formally tests these relationships using our three measures of skill complexity: specialisation, diversity, and frontier position. The results provide strong support for our theoretical framework. Specialisation is strongly associated with sorting into higher wage occupations but only weakly related to adaptive outcomes, confirming that productive depth alone is insufficient for resilience. Diversity, by contrast, is positively associated with skill acquisition, occupational mobility, and transitions into occupations with lower AI exposure, consistent with its interpretation as adaptive capital.

Finally, Hypothesis 3 also receives strong support. Workers closer to the diversity frontier, that is, those who achieve the highest attainable diversity for their level of specialisation, display the strongest adaptive outcomes, particularly skill acquisition, occupational mobility, and AI skill adoption. Frontier position is not associated with stronger sorting into higher wage occupations, reinforcing our argument that productive and adaptive capital capture distinct dimensions of human capital. Overall, the findings support our central claim that worker resilience depends not on maximising specialisation or diversity independently, but on combining specialised expertise with the greatest feasible breadth across knowledge domains.

We provide a more detailed description of the model specifications and regression results in Appendix A4. We also consider an alternative modelling approach in Appendix A5. We dissect the effect of moving closer to the diversity frontier by age, gender, and education in Appendix A6.

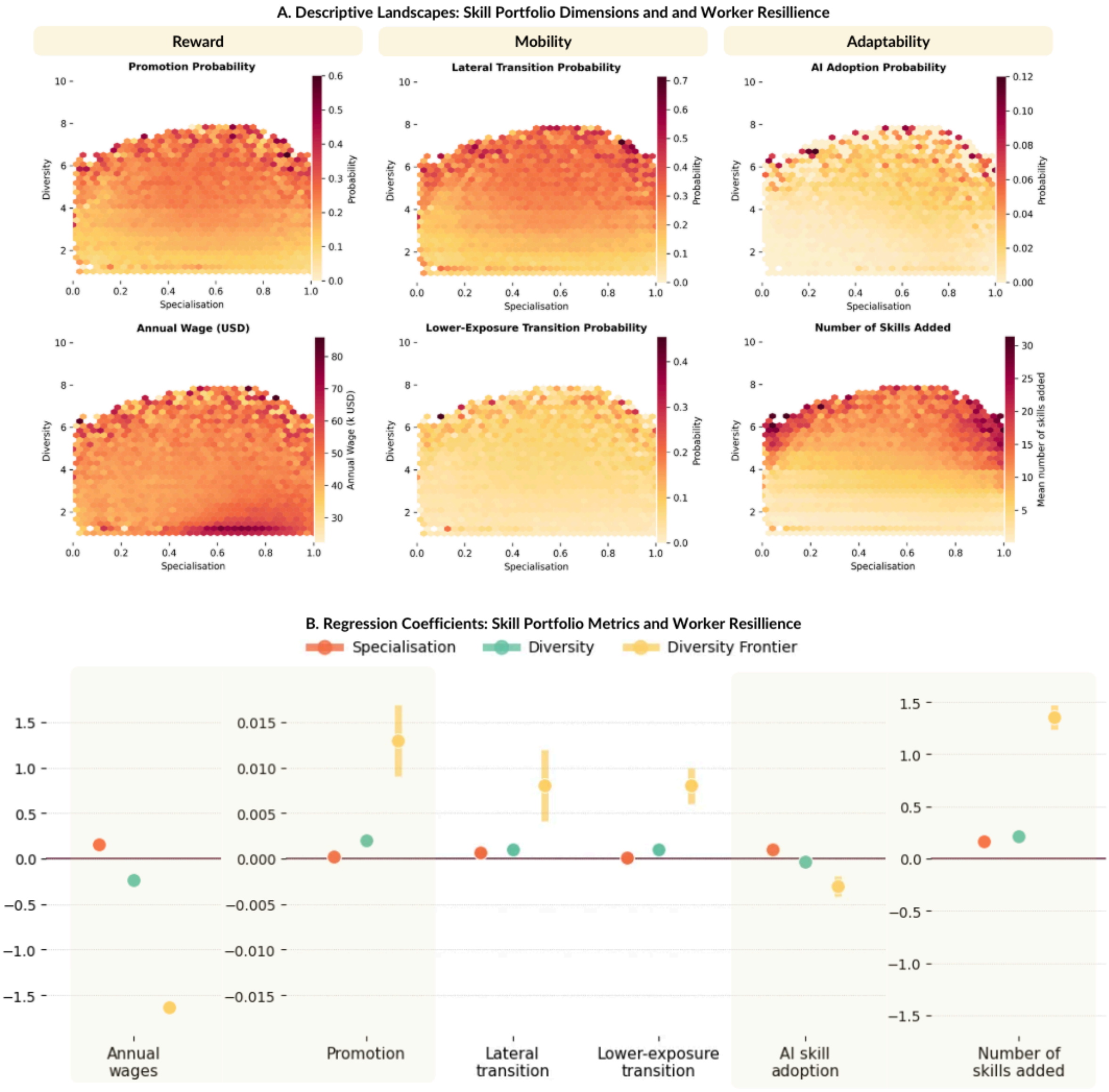

**Figure 3. Skill portfolio dimensions and worker resilience**
Panel A shows descriptive outcome landscapes across workers' specialisation and diversity scores. Darker areas indicate higher outcome probabilities or levels. Reward outcomes are more closely associated with specialisation, while mobility and adaptability outcomes increase more visibly with diversity and, for some outcomes, with the combination of high diversity and high specialisation. Panel B reports regression coefficients for standardised specialisation, standardised diversity, and the standardised diversity frontier score across the main outcomes. The estimates show that specialisation is most strongly associated with occupation-level wage sorting, while diversity and frontier position are more closely associated with mobility and skill acquisition.

# Discussion

This study advances the literature on skill complexity and worker resilience in three ways. First, we introduce an agnostic, data driven representation of workers' skill portfolios. Rather than relying on predefined occupational classifications or expert taxonomies, we recover the structure of the skill space directly from empirical patterns of skill co-occurrence. This allows us to position workers simultaneously according to the depth and breadth of their capabilities and provides a flexible framework for studying how skill portfolios evolve over time.

Second, we distinguish between productive capital and adaptive capital as two complementary dimensions of human capital. Consistent with our theoretical framework, specialised skill portfolios are primarily associated with productive outcomes, particularly sorting into higher wage occupations, whereas diverse portfolios are more strongly associated with adaptive outcomes, including skill acquisition, occupational mobility, transitions into occupations with lower AI exposure, and engagement with emerging AI skills. These findings suggest that labour market value and adaptability represent distinct dimensions of worker resilience.

Third, we introduce the concept of the diversity frontier. Rather than viewing diversity independently of specialisation, we argue that the feasible diversity of a worker's skill portfolio is constrained by their degree of specialisation. Workers closest to the diversity frontier, those who achieve the greatest attainable diversity for their level of specialisation, consistently exhibit the strongest adaptive outcomes. Adaptive capacity therefore depends not simply on possessing broad or specialised skills, but on combining specialised expertise with the broadest feasible set of complementary capabilities.

## Limitations

Several limitations should be considered when interpreting these findings. First, the analysis relies on LinkedIn profiles of United States workers, which over represent digitally engaged and highly skilled occupations. Second, listed skills are self reported and may capture signalling as well as actual capabilities. Third, the construction of the skill network depends on modelling choices, including network pruning, community detection, and the measurement of hierarchy through local reaching centrality. Fourth, the analysis remains associational rather than causal, despite the use of lagged explanatory variables and extensive controls. Finally, the diversity frontier should be interpreted as a relative measure of portfolio structure rather than an independent dimension of human capital. Together, these limitations suggest that the findings should be interpreted as evidence of systematic relationships between skill portfolio structure and subsequent career outcomes rather than causal effects.

## Implications

These findings have implications for workers, firms, and policy makers. For workers, understanding the structure of their skill portfolio may be as important as understanding the individual skills they possess. Our framework highlights not only current capabilities but also complementary and potentially dormant skills accumulated through previous education and employment. Mapping these portfolios can reveal adjacent career opportunities and identify reskilling pathways that build on existing strengths rather than requiring entirely new careers.

For firms, the results suggest that workforce planning should move beyond occupations and job titles toward skill portfolio analysis. Employees who appear similar based on their current role may differ substantially in their adaptive potential because they possess different combinations of specialised and complementary skills. Identifying these hidden capabilities can improve internal mobility, workforce redeployment, succession planning, and targeted learning investments.

For governments and labour market institutions, the framework offers a more granular approach to workforce development. Rather than treating occupations as homogeneous, policy makers could use empirical skill maps to identify transferable capabilities, design personalised reskilling pathways, and support workers in transitioning toward occupations that build on their existing portfolios. Such approaches may become increasingly valuable as technological change accelerates and career transitions become more frequent.

## Outlook

The framework introduced here opens several promising directions for future research. An important next step is to validate these findings using representative labour market surveys that collect detailed information on workers' skills across occupations and career stages. Similar empirical skill networks could be constructed from survey data, allowing the concepts of productive capital, adaptive capital, and the diversity frontier to be examined beyond online professional platforms.

More broadly, our framework provides a foundation for designing evidence based reskilling policies. By identifying workers' existing capabilities and mapping feasible pathways toward adjacent skill portfolios, policy makers and employers could support transitions that build on workers' accumulated human capital rather than replacing it. As technological change increasingly reshapes labour markets, understanding not only the skills workers possess today but also their capacity to acquire complementary skills tomorrow will become central to building an adaptive workforce.

# Appendices

## A1. Theoretical Framework

Formally, we ground our analysis in the framework of economic complexity (Hidalgo and Hausmann, 2009; Tacchella *et al.*, 2012), which conceptualises actors as bundles of capabilities whose diversity and ubiquity jointly determine productive potential. Applied to workers, this perspective treats skills as observable proxies for underlying capabilities, and skill portfolios as the empirical expression of workers' productive and adaptive capacity.

### Workers as Bundles of Capabilities

In Hidalgo and Hausmann (2009), a country's economic complexity reflects both how many distinct products it exports (diversity) and how rare those products are across the world economy (ubiquity). Countries that combine high diversity with low-ubiquity exports hold rare capabilities and achieve high economic complexity. We translate this logic to the individual level: workers with more diverse skill portfolios, composed of skills held by fewer other workers, signal higher accumulated capability.

Let $M_{i,s} \in \{0, 1\}$ be the binary skill matrix, where $M_{i,s} = 1$ if worker $i$ holds skill $s$. Following the notation in Hidalgo and Hausmann we define:

$$Skill\,ubiquity = K_s = \sum_i M_{i,s}$$

$$Worker\,ubiquity = K_i = \sum_s M_{i,s}$$

Worker complexity can be expressed by the recursion:

$$k_{i,0} = K_i,\; k_{s,0} = K, s$$

$$k_{i,n} = \frac{1}{k_{i,0}} \sum_s M_{i,s} k_{s,n-1}$$

$$k_{s,n} = \frac{1}{k_{s,0}} \sum_i M_{i,s} k_{i,n-1}$$

The eigenvector of the normalised matrix $M_{i,s} = \frac{M_{i,s}}{(K_s \cdot K_i)}$. Workers with high complexity hold many skills that are held by few others: they signal rare, accumulated capabilities with high productive potential.

Translating the Hidalgo-Hausmann framework from countries to individuals requires care. Country-level exports are externally verified (customs records), whereas LinkedIn skills are self-reported and may reflect signalling in addition to actual capability. Countries are also far fewer in number (~200) and update their export baskets slowly, whereas workers number in the millions and update their listed skills frequently. As a result, our skill rarity ($K_s$) reflects a mixture of true capability scarcity and self-reporting variation. We interpret skill complexity as a relative positional measure within the LinkedIn-observed skill system, not as a direct measure of capability rarity in the underlying population.

## Specialisation: nestedness and hierarchy in the skills space

Simple ubiquity $K_s$ treats all skills symmetrically. The skill network, however, exhibits a hierarchical dependency structure: some skills are prerequisites for others. To capture this, we follow Hosseinioun et al., 2025, through a directed network $G = (V, E)$, where a directed edge $u \rightarrow v$ encodes that skill $u$ is a prerequisite for skill $v$. We characterize each skill's hierarchical position using local reaching centrality:

$$LRC(s) = \frac{|\{v \in V: \exists \text{ directed path } s \rightarrow v\}|}{|V|}$$

Skills with high LRC are upstream, i.e. they are foundational prerequisites held by many workers (high $K_s$). Skills with low LRC are downstream, i.e. they are endpoint capabilities held by few workers (low $K_s$). LRC thus refines ubiquity by capturing not only how many workers hold a skill, but its structural position as a prerequisite versus an endpoint in the dependency chain.

The specialisation of worker i's portfolio is:

$$Specialisation_i = (1/K_i \sum_s M_{i,s}[1 - LRC_{norm(s)}]) \ ,$$

Which increases as skills concentrate toward rare, downstream endpoints. This operationalises the low-ubiquity dimension explained above, through the hierarchical structure of the skills network.

## Diversity across domains

Diversity across skill domains can be measured using Hill diversity:

$$Diversity_i = exp(- \sum_c p_{i,c} ln(p_{i,c}))$$

where $p_{i,c} = \frac{n_{i,c}}{K_i}$ is the share of a worker i's skills belonging to cluster c, and clusters are defined via community detection on the skill network. $H(s_i)$ equals the effective number of skill domains represented, accounting for both the number of domains and the balance across them.

## A2. Skill Network

We construct a directed skill network based on observed co-occurrence of skills across workers. Starting from a bipartite representation of workers and skills, we compute pairwise co-occurrence counts between skills. The distribution of co-occurrence counts is highly skewed. While the median pair of skills co-occurs in only 108 profiles, a small number of highly prevalent combinations—typically involving general-purpose skills—appear thousands of times, with the top 1% exceeding 3,700 co-occurrences and the maximum reaching over 300,000. This heavy-tailed structure reflects the influence of skill popularity and underscores the need to distinguish statistically meaningful associations from spurious co-occurrence driven by ubiquitous skills.

To identify statistically meaningful relationships, we compare observed co-occurrence to a null model of random association using a hypergeometric distribution. For each pair of skills (i, j), we compute the expected overlap and corresponding variance under independence, and derive a z-score:

$$z_{ij} = \frac{Observed_{ij} - Expected_{ij}}{\sqrt{Var_{ij}}}, \text{ where } Expected_{ij} = \frac{n_i n_j}{N}, \text{ and } Var_{ij} = n_j \frac{n_i}{N}\left(1 - \frac{n_i}{N}\right)\frac{N-n_j}{N-1}.$$

We retain only statistically significant co-occurrences ($z \geq 3.0$), ensuring that edges reflect meaningful relationships rather than popularity effects. The distribution of z-scores is strongly right-skewed, with a median of 12.6 and a long tail of very high values. As expected in large-scale data, a substantial share of links (approximately 74%) exceed the $z \geq 3$ threshold, reflecting high statistical power rather than uniformly strong relationships. We therefore complement statistical filtering with an additional pruning step based on edge weights to recover a more informative and interpretable network structure.

Additional details on the distribution of z-scores, illustrative examples of skill pairs across the distribution, and a systematic sensitivity analysis of alternative thresholds are provided below.

Drawing on prior approaches to identify asymmetric dependencies in skill and capability networks (Neffke and Henning, 2013; Hosseinioun et al., 2025), we infer directionality by comparing conditional probabilities $P(i\,|j)$ and $P(j\,|i)$, where $P(i|j)$ is the probability of observing skill $i$ among workers who list skill $j$, and *vice versa*. Following this logic, directionality captures hierarchical dependency in the skill network. Because workers who hold a specialised skill v are far more likely to also hold its general prerequisite u than the reverse, $P(u|v) > P(v|u)$. We therefore draw the edge from the prerequisite to the dependent capability, u → v, encoding that mastery of u typically precedes acquisition of v. The edge weight $w_{ij} = |P(i|j) - P(j|i)|$ captures the magnitude of this asymmetry, with the direction assigned from the higher-conditional-probability node to the lower. This convention is consistent with the definition of local reaching centrality in the next section, where upstream foundational skills attain high LRC by reaching many downstream dependents.

The distribution of edge weights is highly skewed, with most links exhibiting relatively weak directional differences (median $w = 0.064$, mean $w = 0.11$) and a smaller subset reflecting strong hierarchical relationships. We therefore apply a pruning threshold to remove weak connections ($w \leq 0.05$), eliminating low-signal links while preserving the core dependency structure. This removes approximately 43% of weak edges representing loose co-occurrence rather than meaningful skill relatedness, resulting in a sparse directed network that captures the hierarchical organisation of skill dependencies.

The selection of the edge-weight pruning threshold is documented below, where we evaluate the trade-off between noise reduction and structural interpretability and show that our results are robust to reasonable variations in both thresholds.

## Z-score selection

To construct the directed skill network, we apply a threshold to the pairwise z-scores that measure the strength of co-occurrence between skills. The choice of this threshold involves a trade-off between removing weak or noisy associations and preserving the structural integrity of the network. This section documents the criteria used to select the threshold and demonstrates the robustness of the resulting network.

## Distribution and interpretation of z-scores

The distribution of z-scores is highly dispersed, ranging from -110.46 to 1514.61 (mean = 29.48), indicating substantial variation in the strength of skill co-occurrence. At the extreme upper tail, very large z-scores are typically driven by near-duplicate or highly overlapping skill pairs. For example, pairs such as *Snowflake–Snowflake Cloud* ($z = 1466.79$), *Crisis Counseling–Crisis Counselling* ($z = 1397.82$), and *Desktop Support–Desk Top Support* ($z = 1389.69$) reflect naming variants or tightly coupled concepts that almost always co-occur. Similarly, pairs like *web3.js–Web3* and *Compensation Management–Compensation* capture strong semantic overlap rather than distinct complementary capabilities.

However, high (but not extreme) z-scores often reflect meaningful and substantive complementarities between skills rather than redundancy. For instance, pairs such as *Python–Data Structures* ($z = 230.99$), *Python–Microsoft Azure* ($z = 214.91$), and *Python–JSON* ($z = 159.78$) represent coherent technical bundles that frequently co-occur in practice. These relationships capture strong, domain-relevant associations rather than duplication, indicating that high z-scores can reflect meaningful structure in the skill space.

By contrast, pairs with z-scores close to zero tend to reflect weak or heterogeneous associations. For instance, randomly sampled pairs in the range [-1, 1) include combinations such as *Python–Strategic Consulting* ($z = 0.40$), *Python–Squarespace* ($z = 0.32$), and *Python–Team Facilitation* ($z = -0.10$), which span loosely related or unrelated domains. Even slightly higher ranges (e.g., $z \in [1, 3)$) include broadly defined or weakly related combinations such as *Python–Portuguese* or *Python–Remote Teamwork*. These examples indicate that very low thresholds retain a substantial number of noisy or weakly informative links.

At the lower tail, strongly negative z-scores capture systematically underrepresented combinations. For example, pairs such as *Python–Management* (z = -85.02), *Python–Sales* (z = -96.41), and *Python–Strategic Planning* (z = -71.01) occur far less frequently than expected given their marginal frequencies, reflecting structural separation between technical and managerial skill domains.

Taken together, these patterns suggest that meaningful signal lies between these extremes: thresholds that are too low admit noisy and weak associations, while excessively high thresholds risk removing a large number of substantively meaningful connections, retaining only the strongest and most tightly coupled relationships.

## Stability of network structure across thresholds

To assess the sensitivity of the network to the threshold choice, we reconstruct the network across a range of z-score thresholds and recompute local reaching centrality (LRC) for each specification.

The results show that LRC values are highly stable across thresholds. Figure A2.1 (left) presents the pairwise correlation of LRC values across all thresholds, indicating consistently high similarity, particularly among nearby thresholds. This suggests that the relative ranking of skills in the network is largely invariant to moderate changes in the threshold.

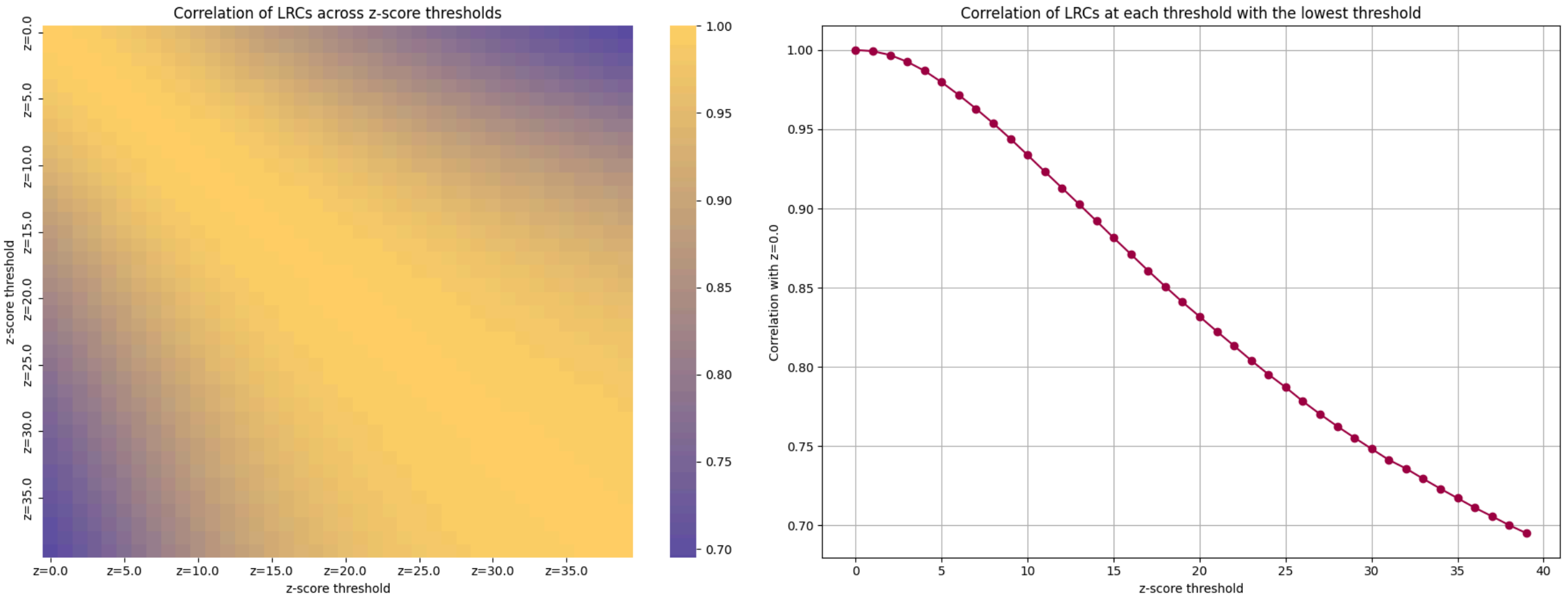

**Figure A2.1**
Left: Pairwise correlation of local reaching centrality (LRC) across z-score thresholds. High correlations across a wide range of thresholds indicate that the ranking of skills by LRC is robust to threshold choice. Right: Correlation of LRC values at each z-score threshold with the baseline specification ($z \geq 0$). The smooth decline indicates gradual changes in network structure rather than discrete shifts.

Complementing this, Figure A2.1 (right) plots the correlation of each threshold with the baseline specification ($z \geq 0$). The relationship exhibits a smooth and gradual decline as the threshold increases, without any sharp discontinuities. This indicates that changes in the network structure occur progressively rather than being driven by any particular cutoff.

At the meso level, community structure is similarly stable. Between z = 1 and z = 10, the number of detected clusters remains remarkably consistent, varying only between approximately 18 and 21 clusters (or 10 and 13 clusters with $\geq$ 30 skills). Figure A2.2 shows that both the total number of clusters and the number of clusters above a minimum size threshold remain stable across this range. This indicates that the modular structure of the network is largely invariant to moderate threshold changes.

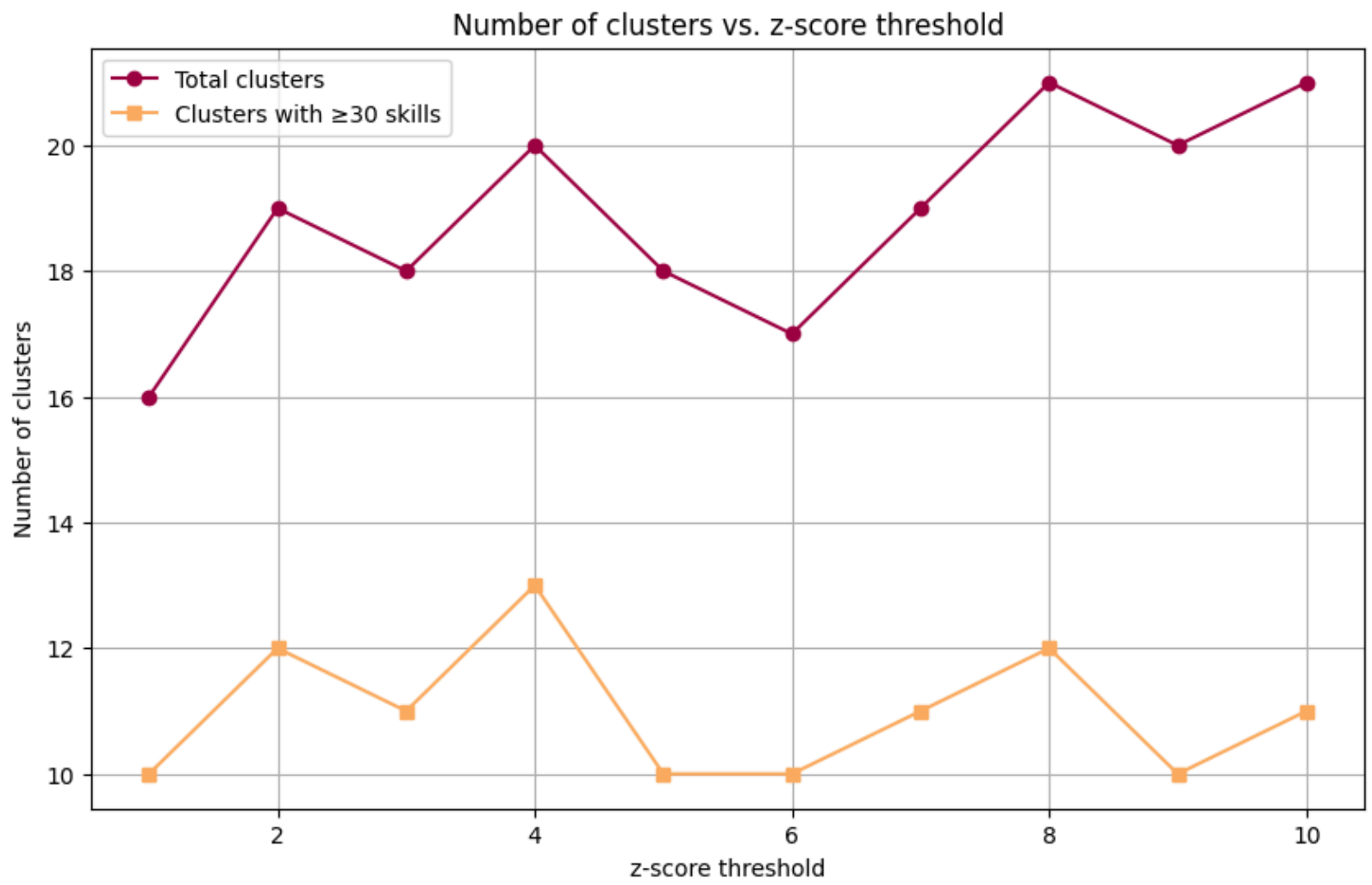

**Figure A2.2**
Number of detected clusters across z-score thresholds. Both total clusters and clusters above a minimum size threshold remain stable across moderate thresholds, indicating robustness of community structure.

## Trade-off between noise reduction and over-pruning

While the overall structure of the network is stable across a wide range of thresholds, increasing the threshold progressively sparsifies the graph. As the threshold rises from z $\geq$ 0 to higher values, the number of edges decreases substantially (e.g., from approximately 1.08 million at z $\geq$ 0 to 0.73 million at z $\geq$ 10), while the number of nodes also declines. At the same time, the number of weakly connected components gradually increases, indicating growing fragmentation of the network.

Higher thresholds also compress the distribution of LRC values, reducing the variation in hierarchical positions across skills. This suggests that excessively strict thresholds begin to remove meaningful connections and flatten the structure of the network. In other words, while higher thresholds reduce noise, they also risk eliminating substantively relevant relationships between skills.

This trade-off is illustrated in Figure A2.3, which shows how the number of nodes, edges, and overall network density evolve as the z-score threshold increases. All three metrics decline smoothly as the threshold becomes more restrictive, indicating progressive sparsification of the network. While this reduction helps eliminate weaker associations, it also reflects the loss of potentially meaningful connections at higher thresholds.

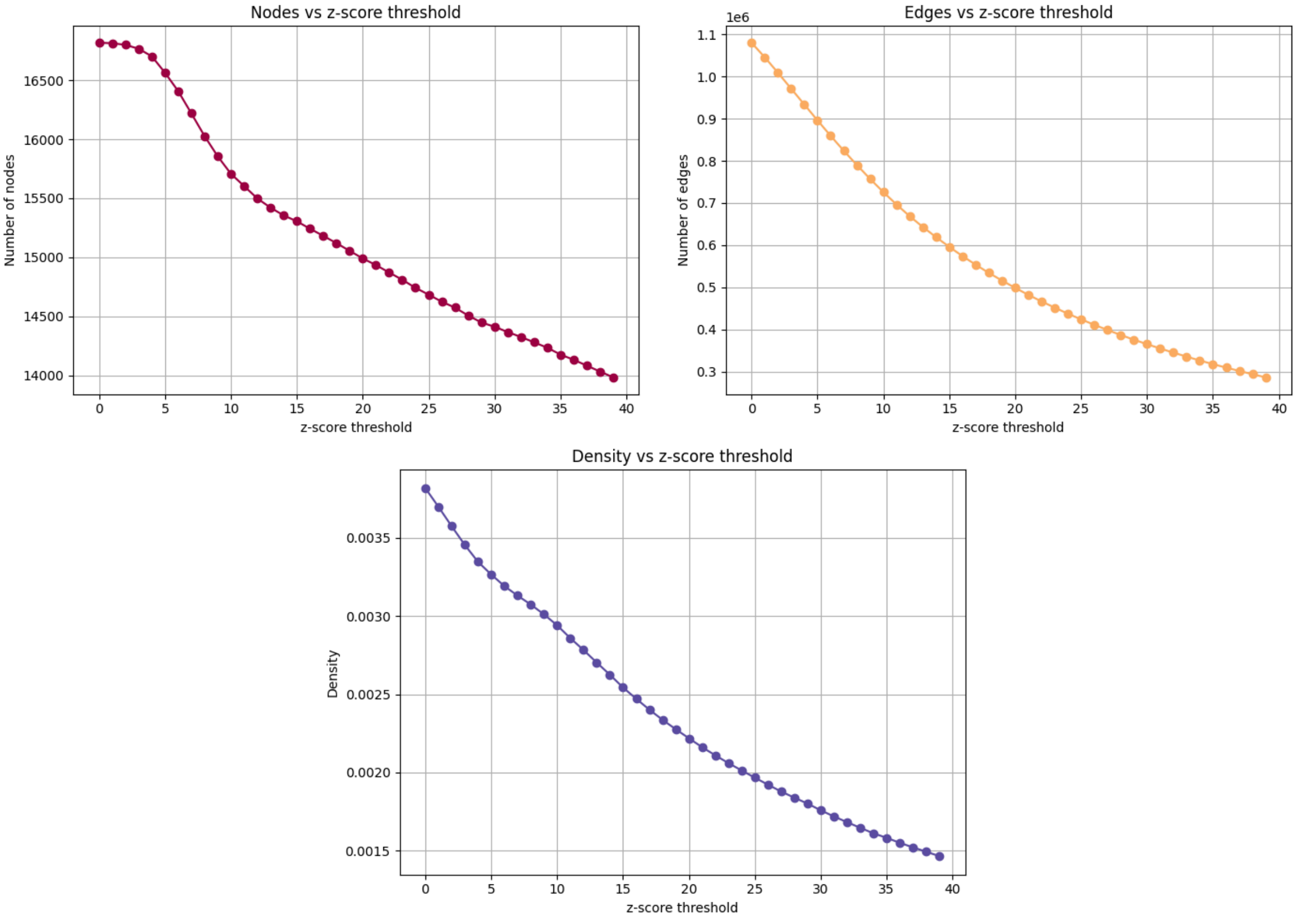

**Figure A2.3**
Evolution of network size and density across z-score thresholds. The number of nodes, edges, and overall network density decline smoothly as the threshold increases, reflecting progressive sparsification of the network. While higher thresholds reduce the number of weak or noisy connections, they also remove a substantial share of edges and nodes, illustrating the trade-off between noise reduction and over-pruning.

## Threshold selection

Based on these results, we select a threshold of $z \geq 3$. This choice reflects a minimal filtering strategy within the region of structural stability, balancing the removal of weak associations with the preservation of network connectivity.

First, it excludes clearly weak and heterogeneous links that are prevalent at very low z-scores, thereby reducing noise in the network. Second, it lies well within the region where both LRC values and community structure are highly stable, ensuring that the results are not sensitive to the exact cutoff. Third, it avoids the over-pruning observed at higher thresholds, which leads to a loss of connectivity and increasing fragmentation. While higher thresholds (e.g., $z \geq 10$) produce similar cluster structures, they remove a substantially larger number of edges and nodes without providing additional structural clarity, indicating diminishing returns from stricter filtering.

Importantly, the stability analysis demonstrates that the substantive results are robust to alternative threshold choices within a broad range (approximately $z = 1$ to $z = 10$). The selected threshold should therefore be interpreted as a conservative filtering choice that preserves meaningful connectivity while removing low-information links, rather than as a parameter that drives the findings.

Overall, this approach ensures that the network construction is guided by empirical structure rather than arbitrary parameter choices, reinforcing the robustness of the subsequent analysis.

## Alpha threshold selection

Following the identification of statistically significant co-occurrences ($z \geq 3$), we apply an additional pruning step based on edge weights, which capture asymmetry in conditional skill relationships. This step removes weak directional dependencies and sharpens the hierarchical structure of the network.

The choice of the pruning threshold ($\alpha$) involves a trade-off between eliminating low-information links and preserving the structural and semantic integrity of the network. This section documents the criteria used to select $\alpha$ and demonstrates the robustness of the resulting network.

## Structural evolution across thresholds

The sensitivity of network structure to the threshold selection is shown in Figure A2.4. Increasing $\alpha$ progressively sparsifies the network. The number of edges declines sharply from approximately 971,000 at $\alpha = 0$ to 556,000 at $\alpha = 0.05$, and below 300,000 at $\alpha \geq 0.125$. Node counts remain largely stable until high thresholds ($\alpha \geq 0.25$). Network density decreases smoothly throughout.

At the same time, the number of detected clusters increases from 11 at $\alpha = 0$ to over 20 at higher thresholds, and the number of weakly connected components rises, indicating growing fragmentation, as seen in Figure A2.5.

This pattern reflects a gradual transition from a dense, highly connected network to a sparser and increasingly partitioned structure.

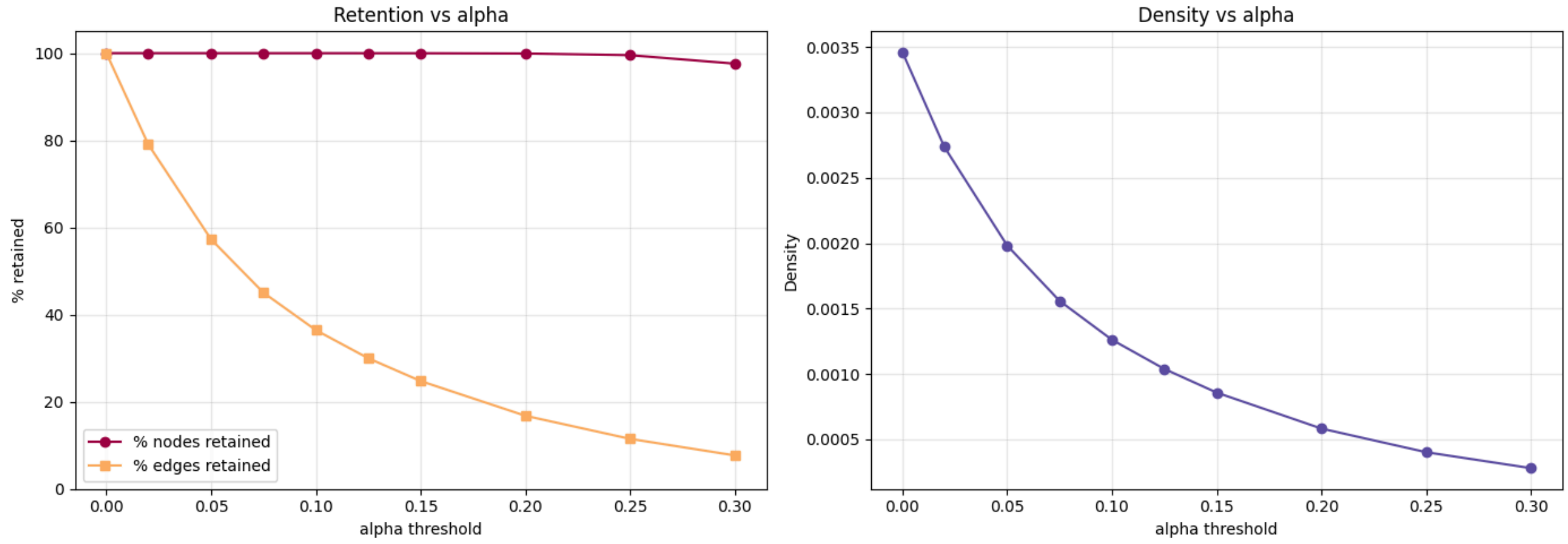

**Figure A2.4. Evolution of network size and density across α thresholds.**

The number of nodes remains largely stable across thresholds, while the number of edges and overall network density decline smoothly as α increases, reflecting progressive sparsification of the network.

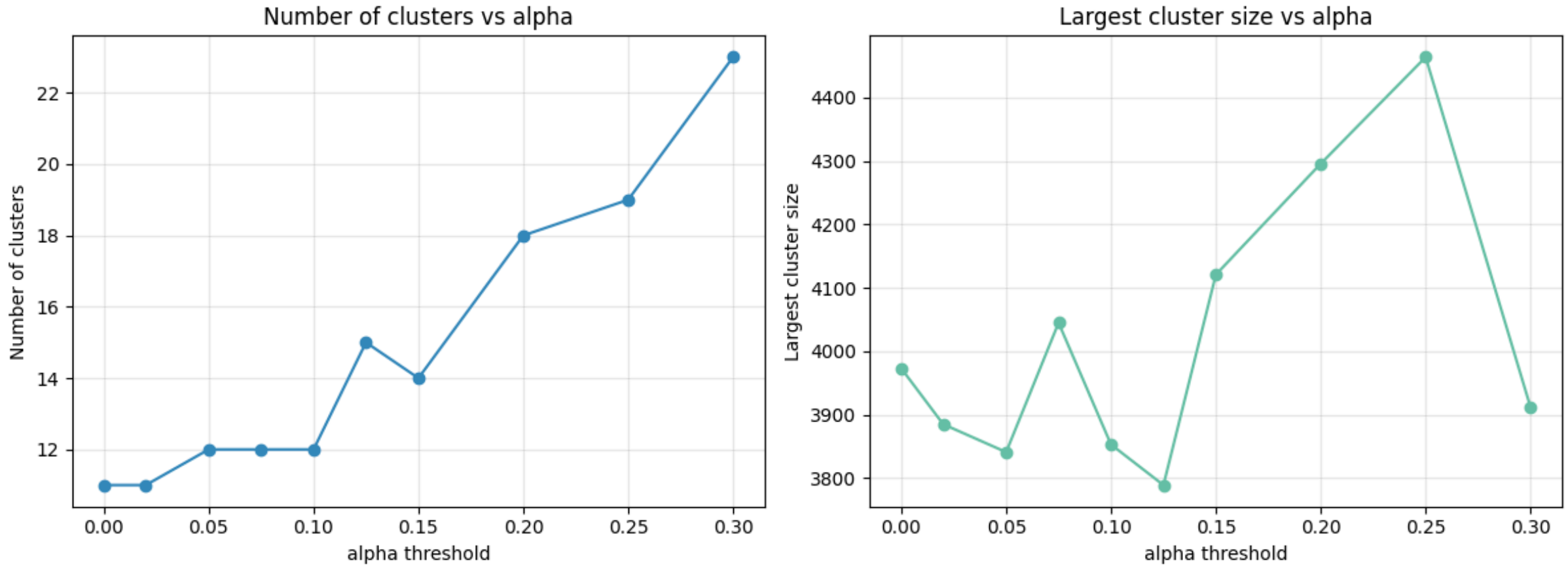

**Figure A2.5. Evolution of community structure across α thresholds.**
The number of detected clusters increases with higher α thresholds, while the size of the largest cluster remains relatively stable at moderate thresholds before declining at higher values, indicating increasing fragmentation of the network.

## Stability of cluster structure

To assess the sensitivity of community structure, we compute Adjusted Rand Index (ARI) and Normalised Mutual Information (NMI) relative to the unpruned baseline ($\alpha = 0$). The results are shown in Figure A2.6.

Clustering similarity metrics (ARI/NMI) provide mixed guidance for threshold selection. In particular, similarity is slightly higher at $\alpha = 0.075$ than at $\alpha = 0.05$, indicating that the resulting partition is more closely aligned with the unpruned baseline. However, these metrics capture resemblance to the baseline partition rather than the interpretability or substantive validity of clusters.

To assess this, we complement the quantitative analysis with a semantic evaluation of cluster composition based on the highest-LRC skills.

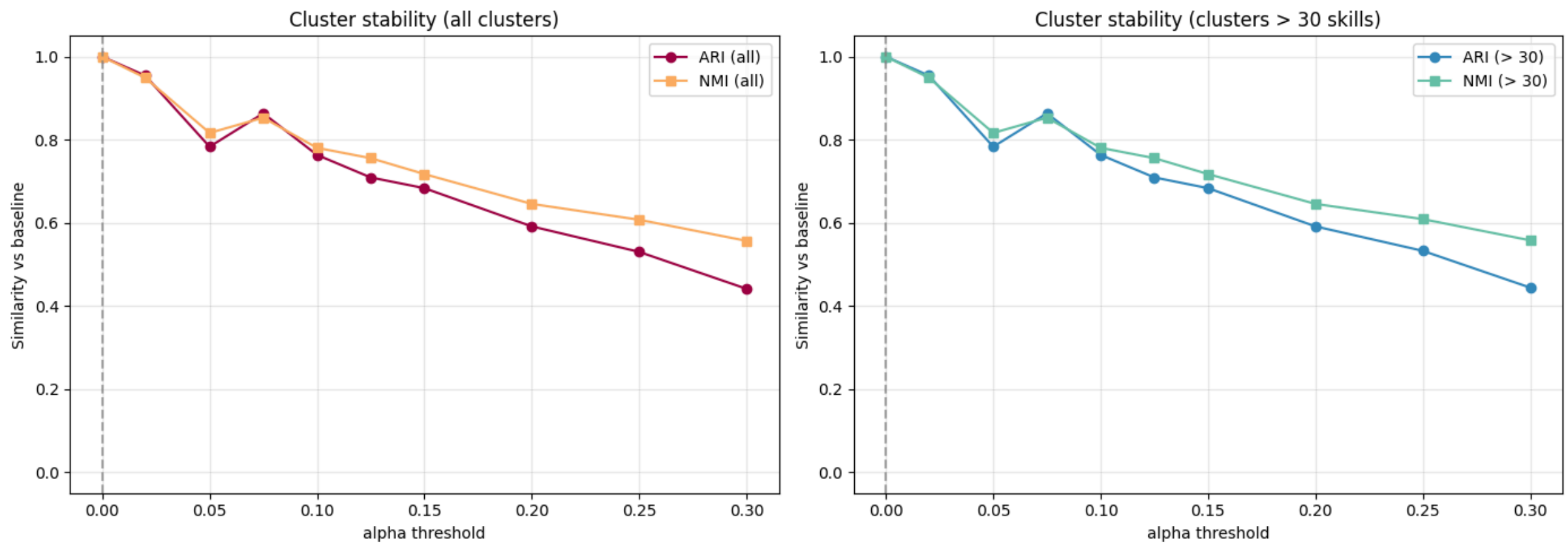

**Figure A2.6. Clustering similarity across α thresholds.**
Adjusted Rand Index (ARI) and Normalised Mutual Information (NMI) are computed relative to the unpruned baseline ($\alpha = 0$). High similarity at low thresholds indicates minimal pruning, while changes at higher thresholds reflect reorganisation of cluster structure. Results are shown for all clusters and for clusters above a minimum size threshold.

## Semantic validation of clusters

To complement these quantitative diagnostics, we examine the semantic composition of clusters using the highest local reaching centrality (LRC) skills within each cluster.

At $\alpha = 0.05$, clusters align closely with broad, recognisable occupational domains, including management and leadership, finance, marketing, legal services, information technology, software and data, healthcare, government and public safety, human resources, and manufacturing. These clusters exhibit high internal semantic coherence and clear separation across domains.

At $\alpha = 0.075$, the same broad domains remain present, but several clusters become subdivided into narrower groupings. For example, leadership and management separate from project and engineering-related management skills, and HR-related activities become more specialised. While this increases granularity, it reduces comparability across domains and leads to a more fragmented representation of the skill space.

At higher thresholds ($\alpha \geq 0.10$), this pattern intensifies, with clusters becoming smaller, more numerous, and increasingly specialised, often reflecting narrow or domain-specific groupings rather than general occupational structures.

## Edge-level diagnostics

To further assess the impact of pruning, we examine edge-loss patterns for Python, a representative high-centrality skill that plausibly exemplifies a hierarchical specialisation pathway.

At $\alpha = 0.05$, pruning removes a limited number (25 edges) of low-weight connections, primarily linking Python to more peripheral or loosely related domains (e.g., IT management, ERP systems, or general organisational skills). In contrast, increasing the threshold to $\alpha = 0.075$ results in a substantial loss of connections (268 edges), including links to applied technical domains (e.g., engineering, cybersecurity, and domain-specific applications). At higher thresholds, edge removal accelerates further, stripping away large portions of the skill's neighbourhood.

This sharp increase in edge loss suggests a non-linear transition between $\alpha = 0.05$ and $\alpha = 0.075$, where pruning moves from removing weak, noisy links to removing substantively meaningful connections that embed skills within broader occupational contexts. As a result, higher thresholds risk artificially narrowing the connectivity of skills and reducing their embeddedness within broader domains.

## Trade-off between noise reduction and over-pruning

Taken together, the results identify three regimes: (i) Low thresholds ($\alpha \leq 0.02$): Minimal pruning; the network remains dense and retains weak or noisy connections; (ii) Intermediate thresholds ($\alpha \approx 0.05$): Substantial reduction of weak edges; clusters become more distinct while preserving broad domain structure; (iii) High thresholds ($\alpha \geq 0.075$): Increasing sparsity and cluster subdivision; meaningful connections are progressively removed and the network becomes more fragmented.

Although clustering similarity (ARI/NMI) is marginally higher at $\alpha = 0.075$ than at $\alpha = 0.05$, this reflects greater similarity to the unpruned baseline rather than improved clustering quality. In contrast, semantic and edge-level diagnostics indicate that $\alpha = 0.05$ better preserves meaningful structure while removing low-information links.

### Threshold selection

Based on these results, we select $\alpha = 0.05$. This threshold: removes a substantial share of weak edges, preserves the main domain-level organisation of the network, and maintains semantically coherent and interpretable clusters, while avoiding the fragmentation observed at higher thresholds.

Importantly, the broad structure of the network is robust across a range of intermediate thresholds (approximately $\alpha = 0.02$ to $0.075$), indicating that the substantive findings are not sensitive to the precise choice of $\alpha$. The selected threshold should therefore be interpreted as a filtering choice that balances noise reduction with structural integrity.

## A3. Skill network cluster statistics

Table A3 reports summary statistics for the 12 skill clusters identified using the Louvain community detection algorithm. For each cluster, we report the number of skills, average degree (mean edges per node), the share of edges linking to other clusters, and summary measures of local reaching centrality (LRC). LRC captures the number of downstream nodes reachable from a given skill in the directed network. To characterise its distribution within clusters, we report the mean, maximum, and Gini coefficient, as well as the share of skills with zero LRC.

Cluster sizes vary substantially, ranging from 242 skills in Legal and Litigation to 3,841 in Administration and Customer Service. Average degree ranges from 16.1 to 50.0 across clusters, indicating variation in local connectivity. The share of edges connecting to other clusters ranges from 18.2% to 54.2%.

The distribution of LRC is highly skewed across all clusters. Maximum values range from 1,095 to 16,716, while mean values are substantially lower, reflecting the presence of a small number of high-centrality nodes. Gini coefficients range from 0.88 to 0.99, indicating a high degree of concentration in LRC within clusters. In addition, a substantial proportion of skills in each cluster have zero LRC, with shares ranging from 45% to 79%. Median LRC is equal to zero in nearly all clusters.

**Table A3. Cluster-level network statistics**
Each row corresponds to a skill cluster identified using the Louvain community detection algorithm. The table reports the number of skills in each cluster, the average local reaching centrality (LRC), the share of edges connecting to other clusters (cross-cluster share), the maximum LRC observed within the cluster, the Gini coefficient of LRC (capturing the dispersion of centrality across skills), and the share of skills with zero LRC. Cross-cluster share and zero LRC share are reported as proportions (0–1).

| Cluster | Skills | Avg. LRC | Cross-cl. share | Max LRC | Gini LRC | Zero LRC share |
|---|---|---|---|---|---|---|
| Leadership, Commercial Management | 2,711 | 241.36 | 0.54 | 16,635 | 0.98 | 0.71 |
| Finance, Accounting, Risk | 759 | 143.72 | 0.44 | 13,425 | 0.96 | 0.62 |
| Marketing, Media, Creative | 1,749 | 175.62 | 0.28 | 15,277 | 0.96 | 0.63 |
| Legal, Litigation | 242 | 46.52 | 0.20 | 1,095 | 0.90 | 0.61 |
| Administration, Customer Service | 3,841 | 134.55 | 0.54 | 16,716 | 0.99 | 0.79 |
| HR, Recruiting, Training | 482 | 193.36 | 0.54 | 15,163 | 0.96 | 0.61 |
| IT Infrastructure, Cybersecurity | 1,022 | 134.52 | 0.32 | 11,229 | 0.96 | 0.64 |
| Government, Military, Public Safety | 273 | 69.87 | 0.28 | 2,166 | 0.88 | 0.45 |
| Healthcare, Clinical Practice | 1,142 | 88.08 | 0.18 | 9,519 | 0.95 | 0.56 |
| Data, AI, Software Engineering | 1,985 | 178.20 | 0.23 | 12,478 | 0.96 | 0.65 |
| Education, Research, Non-profit | 1,489 | 107.47 | 0.46 | 15,541 | 0.97 | 0.60 |
| Engineering, Manufacturing, Construction | 1,058 | 91.78 | 0.36 | 7,132 | 0.95 | 0.59 |

*Note:* Cross-cluster share and zero LRC share are reported as proportions (0–1).

## A4. Model specification and results

Table A4.1 reports the main continuous regression specifications used in the paper. The models use three standardised worker-level skill-portfolio measures: specialisation, measured as average inverse normalised local reaching centrality; diversity, measured as Hill diversity across skill domains; and frontier position, measured by the diversity frontier score. Because each variable is standardised, coefficients can be interpreted as the association between a one-standard-deviation increase in the corresponding portfolio measure and the outcome.

The model sequence introduces each skill complexity measure in parallel fully controlled specifications. These models ask whether productive depth, adaptive breadth, and frontier position are associated with different career outcomes. An additional specification includes the interaction between specialisation and diversity as a robustness check.

Table A4.2 shows complete model build-up for a selected career outcome (promotion). Model 1 includes base controls, topology controls, and fixed effects. Controls include (i) holding multiple jobs, (ii) age, (iii) gender (male/female), and (iv) higher education. Fixed effects include (i) profile activity, (ii) network size, and (iii) number of skills for each worker. Models 2–4, add respectively specialisation, diversity, and the diversity frontier score. The table reports the coefficients and standard errors for all explanatory and control variables across the four model specifications.

**Table A4.1 Main regression results across career outcomes**

**Estimated Regression Coefficients (Standard Errors in Parentheses)**

| | | Model Specification | | |
|---|---|---|---|---|
| **Outcome** | **Variable** | **Specialisation (M2)** | **Diversity (M3)** | **Diversity Frontier (M4)** |
| **Promotion** | | | | |
| | 1-LRC Mean | 0.000200<br>(0.000245) | | |
| | $\text{Hill}_1$ | | 0.002***<br>(0.000291) | |
| | Diversity Frontier | | | 0.013***<br>(0.002) |
| **Annual Wages** | | | | |
| | 1-LRC Mean | 0.160***<br>(0.004) | | |
| | $\text{Hill}_1$ | | -0.241***<br>(0.005) | |
| | Diversity Frontier | | | -1.629***<br>(0.032) |
| **Lateral Transition** | | | | |
| | 1-LRC Mean | 0.000672**<br>(0.000243) | | |
| | $\text{Hill}_1$ | | 0.001***<br>(0.000287) | |
| | Diversity Frontier | | | 0.008***<br>(0.002) |
| **Lower-Exposure Transition** | | | | |
| | 1-LRC Mean | 0.000091<br>(0.000149) | | |
| | $\text{Hill}_1$ | | 0.001***<br>(0.000176) | |
| | Diversity Frontier | | | 0.008***<br>(0.001) |
| **AI-Skill Adoption** | | | | |
| | 1-LRC Mean | 0.000994***<br>(0.000058) | | |
| | $\text{Hill}_1$ | | -0.000341***<br>(0.000086) | |
| | Diversity Frontier | | | -0.003***<br>(0.000578) |
| **Number of Skills Added** | | | | |
| | 1-LRC Mean | 0.164***<br>(0.007) | | |
| | $\text{Hill}_1$ | | 0.217***<br>(0.009) | |
| | Diversity Frontier | | | 1.355***<br>(0.061) |

*Notes:* Standard errors are reported in parentheses. The Specialisation column reports the individual specialisation specification. The Diversity column reports the individual diversity specification. The Diversity Frontier column reports the individual frontier specification. Specialisation is measured using standardised 1-LRC Mean at $t-1$; diversity is measured using standardised $\text{Hill}_1$ at $t-1$; diversity frontier is measured using the Diversity Frontier Score at $t-1$. All models control for multiple job holder status, profile activity (log standardised), network size (log standardised), age (standardised), gender, and number of skills (standardised), all measured at $t-1$. Models additionally include company size, functional area, and industry fixed effects. N=1,604,114 across all model specifications. $^{*}p < 0.05$; $^{**}p < 0.01$; $^{***}p < 0.001$.

**Table A4.2. Regression model build-up for selected outcome (Promotion)**

**Estimated Regression Coefficients (Standard Errors in Parentheses)**

| Variable | Promotion | | | |
|---|---|---|---|---|
| | **Model 1** | **Model 2** | **Model 3** | **Model 4** |
| **Skill portfolio variables** | | | | |
| $\text{Specialist}_{i,t-1}$ | – | -0.004*** | – | – |
| | | (0.000584) | | |
| $\text{Diverse}_{i,t-1}$ | – | – | 0.003*** | – |
| | | | (0.000653) | |
| Diversity Frontier $\text{Score}_{i,t-1}$ | – | – | – | 0.013*** |
| | | | | (0.002) |
| **Base controls** | | | | |
| Multiple Job $\text{Holder}_{i,t-1}$ | 0.369*** | 0.369*** | 0.369*** | 0.369*** |
| | (0.000766) | (0.000766) | (0.000766) | (0.000766) |
| Age, $\text{std.}_{i,t-1}$ | -0.037*** | -0.037*** | -0.037*** | -0.037*** |
| | (0.000239) | (0.000239) | (0.000239) | (0.000239) |
| Male | -0.000719 | -0.000564 | -0.000682 | -0.000659 |
| | (0.000569) | (0.000570) | (0.000569) | (0.000569) |
| Education is $\text{BA+}_{i,t-1}$ | 0.020*** | 0.020*** | 0.020*** | 0.020*** |
| | (0.000529) | (0.000529) | (0.000529) | (0.000529) |
| **Topology and activity controls** | | | | |
| Profile Activity, log $\text{std.}_{i,t-1}$ | 0.014*** | 0.013*** | 0.013*** | 0.013*** |
| | (0.000462) | (0.000463) | (0.000463) | (0.000463) |
| Network Size, log $\text{std.}_{i,t-1}$ | 0.027*** | 0.027*** | 0.027*** | 0.027*** |
| | (0.000282) | (0.000282) | (0.000282) | (0.000282) |
| Number of skills, $\text{std.}_{i,t-1}$ | -0.003*** | -0.003*** | -0.003*** | -0.003*** |
| | (0.000333) | (0.000333) | (0.000344) | (0.000347) |
| Company size fixed effects | ✓ | ✓ | ✓ | ✓ |
| Functional area fixed effects | ✓ | ✓ | ✓ | ✓ |
| Industry fixed effects | ✓ | ✓ | ✓ | ✓ |
| Observations | 1,604,114 | 1,604,114 | 1,604,114 | 1,604,114 |
| $R^2$ | 0.264 | 0.264 | 0.264 | 0.264 |

*Notes:* Standard errors are reported in parentheses. Model 1 reports the baseline specification with controls and fixed effects. Model 2 replaces the focal skill-portfolio measure with the Specialist indicator. Model 3 replaces it with the Diverse indicator. Model 4 replaces it with the Diversity Frontier Score. Fixed effects are indicated with check marks rather than reported as coefficients. Stars are computed from the displayed coefficients and standard errors using a normal approximation. $^{*}p < 0.05$; $^{**}p < 0.01$; $^{***}p < 0.001$.

## A5. Categorical model specification and results

The main analysis uses continuous measures of specialisation and diversity. As a robustness check, Table A5 and Figure A5 examine whether the main patterns are also visible when workers are classified using threshold-based definitions. In these specifications, workers are classified as specialists if their specialisation score is above a given percentile of the specialisation distribution and as diverse if their domain-diversity score is above a given percentile of the diversity distribution.

Figure A5 reports cutoff robustness estimates across alternative top-percentile thresholds from 10 to 90 percent. Each point shows the estimated coefficient from a regression in which the treatment indicator equals one for workers above the corresponding cutoff. The shaded areas report 95 percent confidence intervals. The blue line shows estimates for the specialist-worker definition, while the orange line shows estimates for the diverse-worker definition.

The purpose of this exercise is not to select an optimal cutoff for the main analysis, since the preferred specification uses continuous measures. Instead, the figure evaluates whether the substantive conclusions depend on a particular threshold-based definition. The results indicate that they do not. Across thresholds, the broad patterns are consistent with the main continuous specifications: specialisation is most strongly associated with sorting into higher-wage occupations, while diversity is more consistently associated with mobility and skill acquisition. Coefficients vary in magnitude across cutoffs, as expected, because stricter and looser thresholds compare different parts of the skill-portfolio distribution. However, the qualitative distinction between specialisation as a reward-related measure and diversity as an adaptability-related measure remains visible across the distribution.

Overall, the cutoff robustness exercise supports the interpretation that the main findings are not an artefact of modelling specialisation and diversity as continuous variables. Similar patterns are visible when the measures are converted into substantively interpretable high-specialisation and high-diversity groups.

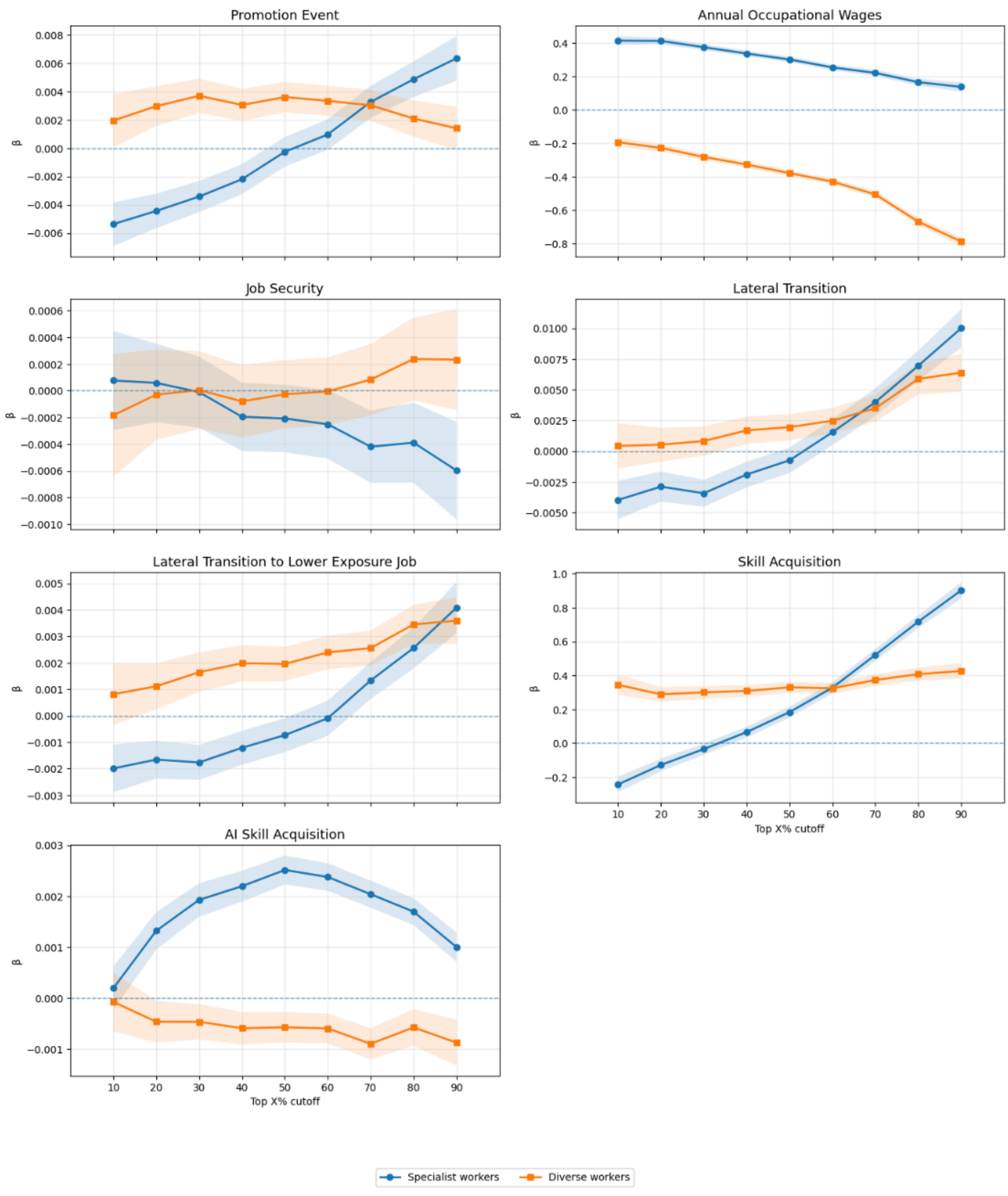

**Figure A5. Cutoff robustness for specialist and diverse worker definitions**
The figure reports coefficient estimates from single-indicator models that redefine specialist and diverse workers using alternative top-percentile cutoffs from 10 to 90 percent. The blue line shows estimates for specialist workers, and the orange line shows estimates for diverse workers; shaded areas report 95 percent confidence intervals. The results show that the main patterns are not driven by the top-quartile threshold: specialisation remains most strongly associated with occupational wages and AI-skill adoption, while diversity is more consistently associated with mobility and skill acquisition. Coefficients vary in magnitude across cutoffs, but the broad distinction between specialisation as a reward-related measure and diversity as an adaptability-related measure remains visible across the distribution.

## A6. Diversity frontier score

To assess whether diversity is associated with adaptive outcomes among workers with comparable levels of specialisation, we construct a diversity frontier score. Workers are first ranked by specialisation and partitioned into 100 equally sized groups. Within each specialisation group, we divide each worker's Hill diversity score by the maximum Hill diversity observed in that group. The resulting measure ranges from 0 to 1 and captures how close a worker is to the observed diversity frontier among workers with similar levels of specialisation. Higher values indicate greater cross-domain diversity relative to similarly specialised workers.

Figure A6.1 plots binned averages of six career outcomes across the diversity frontier score. The figure shows a clear positive gradient for adaptive outcomes: lateral transitions, transitions into lower-exposure roles, total skill acquisition, and AI-skill adoption all increase as workers move closer to the diversity frontier. Promotion also rises with frontier position, suggesting that workers closer to the frontier may be more likely to experience upward mobility as well as lateral mobility. By contrast, occupation-level wages decline sharply at low levels of the frontier score and then flatten, indicating that frontier position is not associated with sorting into higher-wage occupations in the same way as specialisation. These descriptive patterns support the main interpretation that frontier position is more closely associated with learning and mobility than with occupation-level wage sorting.

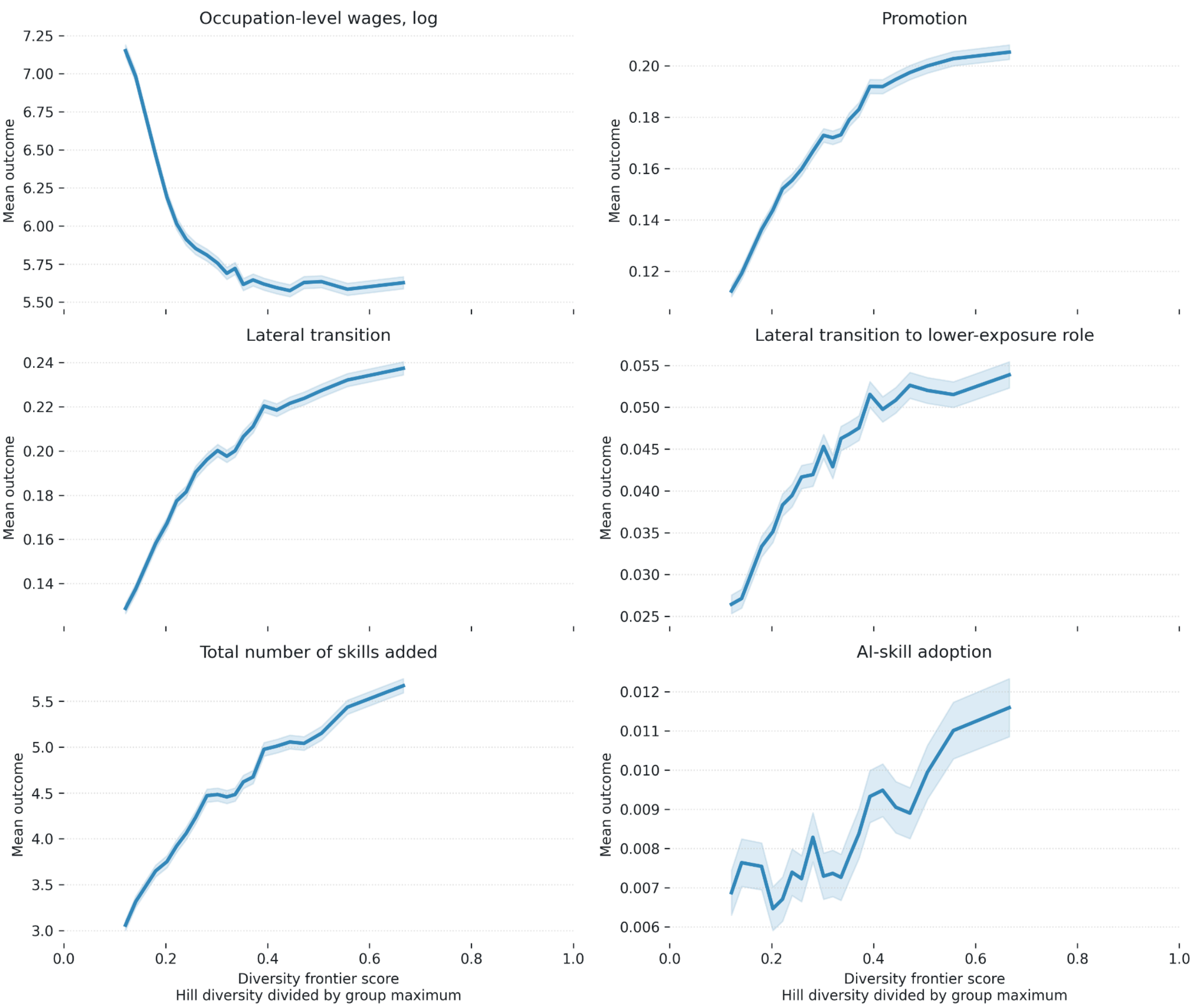

**Figure A6.1 Diversity frontier score and adaptive outcomes**
The figure reports binned averages of six career outcomes across the diversity frontier score: lateral transitions, transitions into lower-exposure roles, total skill acquisition, AI-skill adoption, promotion, and occupation-level wages. Workers are first ranked by specialisation and divided into 100 groups. Within each group, Hill diversity is normalised by the maximum observed Hill diversity. Higher values indicate workers who are closer to the observed diversity frontier among similarly specialised workers. Shaded areas report 95 percent confidence intervals.

Figures A6.2–A6.4 replicate the diversity frontier analysis by gender, age generation, and education. Across groups, the relationship between frontier position and adaptive outcomes remains broadly positive: workers closer to the diversity frontier are more likely to make lateral transitions, transition into lower-exposure roles, add skills, and adopt AI skills. The magnitude of these gradients varies across demographic groups, but the overall pattern suggests that the association between frontier position and adaptive outcomes is not driven by a single subgroup. The subgroup figures also show that promotion and occupation-level wages are less consistently related to frontier position, reinforcing the distinction between adaptive capacity and traditional reward outcomes.

Diversity Frontier Score and Outcomes: By Gender

**Figure A6.2 Diversity frontier score and adaptive outcomes by gender**
The figure reports binned averages of six career outcomes across the diversity frontier score, separately for men and women. Workers are ranked by specialisation and divided into 100 groups; within each group, Hill diversity is normalised by the maximum observed Hill diversity. Higher values indicate workers closer to the observed diversity frontier among similarly specialised workers. Shaded areas report 95 percent confidence intervals.

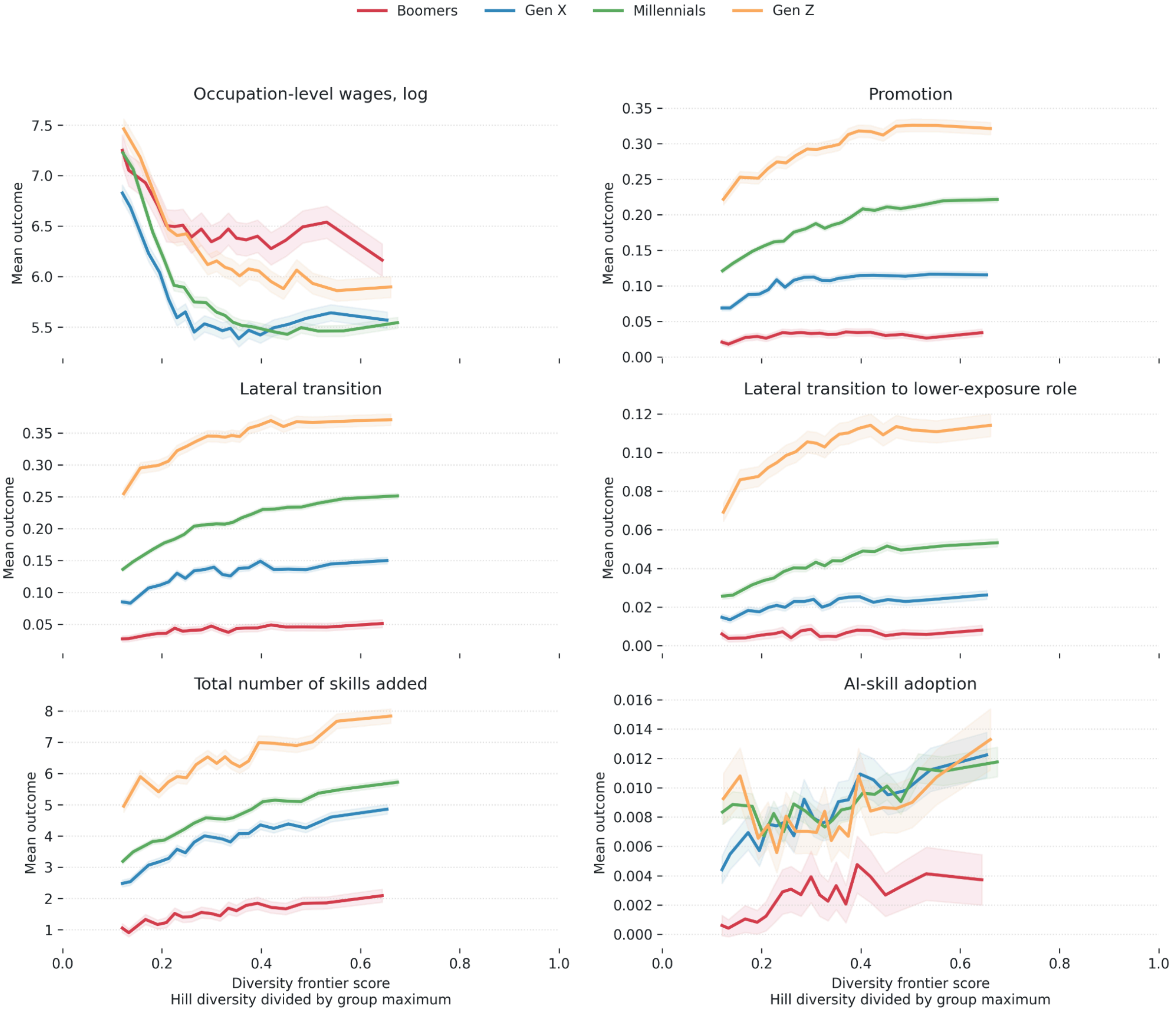

**Figure A6.3 Diversity frontier score and adaptive outcomes by age generation**
The figure reports binned averages of six career outcomes across the diversity frontier score, separately by age generation. Workers are ranked by specialisation and divided into 100 groups; within each group, Hill diversity is normalised by the maximum observed Hill diversity. Higher values indicate workers closer to the observed diversity frontier among similarly specialised workers. Shaded areas report 95 percent confidence intervals.

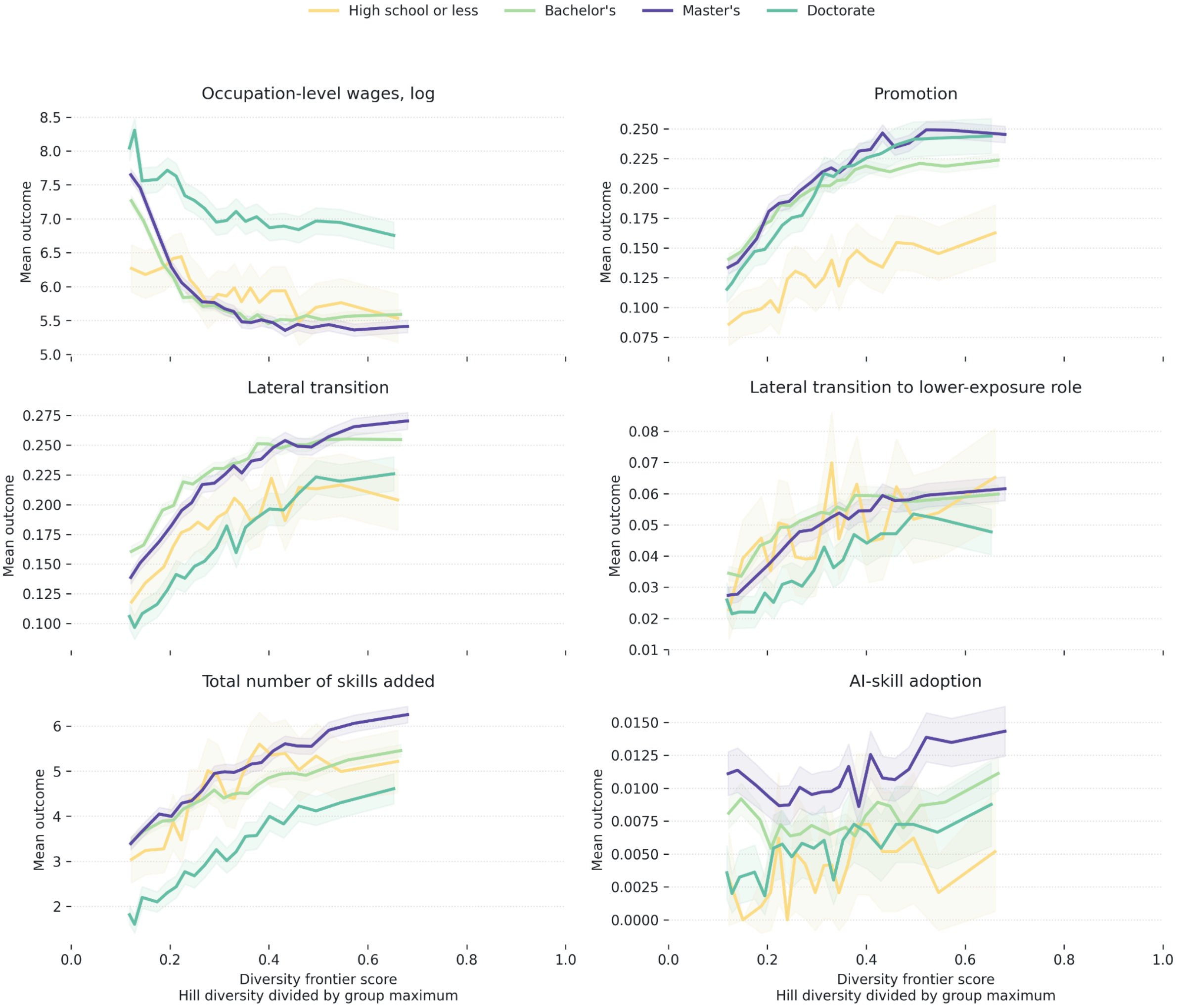

**Figure A6.4 Diversity frontier score and adaptive outcomes by education**
The figure reports binned averages of six career outcomes across the diversity frontier score, separately by educational attainment. Workers are ranked by specialisation and divided into 100 groups; within each group, Hill diversity is normalised by the maximum observed Hill diversity. Higher values indicate workers closer to the observed diversity frontier among similarly specialised workers. Shaded areas report 95 percent confidence intervals.